\documentclass[12pt]{article} 
\usepackage{epsfig, amsmath, amssymb}
\usepackage{hyperref}
\hypersetup{bookmarksopen=true, bookmarksnumbered=true,
 pdfstartview={FitH}, pdfborder={0 0 0}, colorlinks=true,
 linkcolor=red, linktocpage=true, citecolor=blue, urlcolor=blue,
 unicode=true, pdftitle={Entangled Spins and Ghost Spins},
 pdfauthor={DJ,KN}, pdfsubject={Entanglement of positive and
   indefinite norm sectors}, pdfkeywords={dS, EE, b-c ghost system,
   toy model}}
\usepackage{cite}
\newcommand\arXivid[1] {\href{http://arxiv.org/abs/#1}{\tt arXiv:#1}}

\newcommand\atmp[3] {{\it Adv.\ Theor.\ Math.\
    Phys.\ }\href{http://inspirehep.net/search?ln=en&ln=en&p=find+j+''Adv.Theor.Math.Phys.,#1,#3''&of=hb&action_search=Search&sf=&so=d&rm=&rg=25&sc=0}{{\bf #1} (#2) #3}}

\newcommand\cqg[3] {{\it Class.\ Quant.\ Grav.\ }\href{http://inspirehep.net/search?ln=en&ln=en&p=find+j+''Class.Quant.Grav.,#1,#3''&of=hb&action_search=Search&sf=&so=d&rm=&rg=25&sc=0}{{\bf #1} (#2) #3}}

\newcommand\jhep[3]{{\it JHEP\ }\href{http://inspirehep.net/search?ln=en&ln=en&p=find+j+''JHEP,#1,#3''&of=hb&action_search=Search&sf=&so=d&rm=&rg=25&sc=0}
{{\bf #1} (#2) #3}}

\newcommand\npb[4] {{\it Nucl.\ Phys.\ }\href{http://inspirehep.net/search?ln=en&ln=en&p=j+nucl.phys.+#1#2,+#4&of=hb&action_search=Search&sf=&so=d&rm=&rg=25&sc=0}{{\bf #1 #2} (#3) #4}}

\newcommand\pr[4] {{\it Phys.\ Rev.\
  }\href{http://dx.doi.org/10.1103/PhysRev#1.#2.#4}{{\bf #1#2} (#3) #4}} 

\newcommand\pl[4] {{\it Phys.\ Lett.\
  }\href{http://inspirehep.net/search?ln=en&ln=en&p=find+j+\%22Phys.+Lett.,#1#2,#4\%22&of=hb&action_search=Search&sf=&so=d&rm=&rg=25&sc=0}{{\bf
      #1#2} (#3) #4}}

\newcommand\phyrep[3] {{\it Phys.\
    Rept.\
  }\href{http://inspirehep.net/search?ln=en&ln=en&p=find+j+\%22Phys.+Rep.,#1,#3\%22&of=hb&action_search=Search&sf=&so=d&rm=&rg=25&sc=0}{{\bf
      #1} (#2) #3}}

\newcommand\prl[3] {{\it Phys.\ Rev.\ Lett.\
  }\href{http://dx.doi.org/10.1103/PhysRevLett.#1.#3}{{\bf #1} (#2) #3}}

\newcommand\rmp[3]{{\it Rev.\ Mod.\ Phys.\
  }\href{http://dx.doi.org/10.1103/RevModPhys.#1.#3}{{\bf #1} (#2) #3}}

\newcommand{\be}{\begin{equation}}
\newcommand{\ee}{\end{equation}}
\newcommand{\ben}{\begin{equation}}
\newcommand{\een}{\end{equation}}
\newcommand{\bea}{\begin{eqnarray}}
\newcommand{\eea}{\end{eqnarray}}
\newcommand{\bA}{\begin{array}}
\newcommand{\eA}{\end{array}}
\newcommand{\bc}{\begin{center}}
\newcommand{\ec}{\end{center}}
\newcommand{\al}{\alpha}

\newcommand{\ra}{\rightarrow}
\newcommand{\del}{\partial}

\newcommand{\ie}{{\it i.e.}}
\newcommand{\eg}{{\it e.g.}}

\numberwithin{equation}{section}

\begin{document}


\begin{titlepage}
\vspace{25mm}

\bc

\hfill 
\\         [25mm]

{\Huge Entangled spins and ghost-spins}
\vspace{10mm}

{\large Dileep P.~Jatkar$^{1,2}$ and K.~Narayan$^3$} \\
\vspace{3mm}
{\it 1. Harish-Chandra Research Institute \\
Chhatnag Road, Jhusi, Allahabad 211019, India\\[2mm]
2. Homi Bhabha National Institute\\
Training School Complex, Anushakti Nagar, Mumbai 400085, India\\[2mm]
3. Chennai Mathematical Institute, \\
SIPCOT IT Park, Siruseri 603103, India.\\[4mm]}
\ec
\vspace{20mm}

\begin{abstract}
We study patterns of quantum entanglement in systems of spins and
ghost-spins regarding them as simple quantum mechanical toy models for
theories containing negative norm states. We define a single
ghost-spin as in \arXivid{1602.06505} [hep-th] as a 2-state spin
variable with an indefinite inner product in the state space. We find
that whenever the spin sector is disentangled from the ghost-spin
sector (both of which could be entangled within themselves), the
reduced density matrix obtained by tracing over all the ghost-spins
gives rise to positive entanglement entropy for positive norm states,
while negative norm states have an entanglement entropy with a
negative real part and a constant imaginary part. However when the
spins are entangled with the ghost-spins, there are new entanglement
patterns in general.  For systems where the number of ghost-spins is
even, it is possible to find subsectors of the Hilbert space where
positive norm states always lead to positive entanglement entropy
after tracing over the ghost-spins.  With an odd number of ghost-spins
however, we find that there always exist positive norm states with
negative real part for entanglement entropy after tracing over the
ghost-spins.
\end{abstract}

\end{titlepage}


\newpage 
{\footnotesize
\begin{tableofcontents}
\end{tableofcontents}}

\vspace{5mm}

\section{Introduction} \label{sec:introduction}

The concept of entanglement, in some sense, is at the heart of the
interpretation of quantum mechanics.  The entanglement entropy is a
measure of entanglement of two subsystems of a quantum mechanical
system.  Initially, systems with a finite number of degrees of freedom
were investigated by decomposing them into two disjoint subsets and
computing entanglement between these subsets using measures like the
von Neumann entropy or the Renyi entropy. Over the last several years,
these methods have been extended to computing entanglement entropy in
quantum field theories as well. Although the problem of computing
entanglement entropy is substantially more complicated in 
quantum field theories, various techniques have been developed to
evaluate it in specific cases. A partial list of references and
reviews is \cite{AreaLaw1,AreaLaw2,Holzhey:1994we,Vidal:2002rm,
Latorre:2003kg,Calabrese:2004eu,Horodecki:2009zz,Eisert:2008ur,
Calabrese:2009qy,Casini:2009sr}.  In the context of holography
\cite{Maldacena:1997re,Gubser:1998bc, Witten:1998qj,Aharony:1999ti},
the Ryu-Takayanagi formulation
\cite{Ryu:2006bv,Ryu:2006ef,HRT,HEEreview,HEEreview2} via bulk
extremal surfaces has enabled new investigations and perspectives on
quantum entanglement in strongly interacting quantum field theories.

In quantum field theories with a gauge symmetry, one naturally
encounters degrees of freedom which have negative norm.  Although the
physical subspace of the Hilbert space has definite norm, in many
gauge choices we end up having to deal with degrees of freedom with
indefinite norm.  In order to ask questions about entanglement in the
gauge theories, it would be desirable to have a better understanding
of systems which have indefinite norm.  Instead of directly attempting
this in the gauge theory context, it is easier to look at toy models
which are simpler to deal with but at the same time reflect the
intricacies of the system with indefinite norm.  In this paper, we
will consider a system consisting of ordinary spins, which mimic the
positive norm part of the Hilbert space, and ``ghost-spins'' (as
defined in \cite{Narayan:2016xwq}), which incorporate the indefinite
norm sector. Although in the gauge theory context a certain restricted
class of entangled states with mixing of definite and indefinite norm
occur, the toy models distilled from them exhibit many interesting
possibilities.  Restricting to the physical Hilbert space then
corresponds to tracing over all the ghost-spins (regarded as invisible) 
in these toy models and looking at the reduced density matrix of 
the spin system.

The motivation for defining ``ghost-spins'' in \cite{Narayan:2016xwq}
(where entanglement entropy in certain ghost CFTs was studied) arose
from $dS/CFT$: we will review this in the Discussion section (sec.~7).
Here we simply explore patterns of quantum entanglement that emerge in
systems containing entangled spins and ghost-spins, regarding them as
toy models for subsectors with negative norm states arising in
covariant formulations of theories with gauge symmetry, as mentioned
earlier.  As we will review in sec.~2, while a single spin is defined
as a 2-state spin variable with a positive definite inner product
$\langle \uparrow|\uparrow\rangle = 1 = \langle \downarrow|\downarrow\rangle$ 
and 
$\langle \uparrow|\downarrow\rangle = 0 = \langle \downarrow|\uparrow\rangle$,
a single ghost-spin is defined, as in \cite{Narayan:2016xwq}, as a 
2-state spin variable with the indefinite inner product
$\langle \uparrow|\uparrow\rangle = 0 = \langle \downarrow|\downarrow\rangle$ 
and 
$\langle \uparrow|\downarrow\rangle = 1 = \langle \downarrow|\uparrow\rangle$,\
akin to the inner products in the $bc$-ghost system as is well-known 
(see \eg\ \cite{polchinskiTextBk}).
For multiple variables, the spin Hilbert space has a positive
definite metric $g_{ij}=\delta_{ij}$, while the ghost-spin states have
a non-positive metric $\gamma_{ij}$, with components\ $\gamma_{++}=1,
\ \gamma_{--}=-1$, by a basis change\ 
$\{|\uparrow\rangle,|\downarrow\rangle\}\ra \{|+\rangle,|-\rangle\}$\ 
which makes negative norm states manifest.  Overall these systems of
spins and ghost-spins appear to give a broad class of toy models with
a lot of flexibility to engineer a variety of quantum mechanical
systems containing negative norm states.

In sec.~\ref{sec:revi-spins-ghost}, we will briefly review the spin
and the ghost-spin system, and patterns of entanglement in the two
ghost-spin system studied in \cite{Narayan:2016xwq}. In
sec.~\ref{sec:rdm-gs}, we define the reduced density matrix $\rho^s_A$
for the remaining spin variables after tracing over all the
ghost-spins by requiring that the correlation function 
$\langle\psi|O_s|\psi\rangle$ (appearing in the expectation value) 
in any given state $|\psi\rangle$ for any observable $O_s$ of spin 
variables alone is identical to that calculated using the density 
matrix of the mixed state of the remaining spins as 
${\rm tr}_s(O_s\rho^s_A)$. In general, the Hilbert space of spins and
ghost-spins contains positive as well as negative norm states. One
might ask if the entanglement entropy $S_A$ of $\rho^s_A$ is uniformly
positive for all positive norm states, and uniformly negative for all
negative norm states.  This can be shown to be identically true
(sec.~\ref{sec:spins-disent-ghost}), when the spin sector is not
entangled with the ghost-spin sector (both of which could be entangled
within themselves).  In this case the state is a product state
comprising spins disentangled from ghost-spins: the sign of the norm
of the state enters as an overall sign in $\rho^s_A$, giving $S_A>0$
for positive norm states, while for negative norm states, $S_A$ has a
negative real part and a constant imaginary part. This is similar to
the case of two ghost-spins studied in \cite{Narayan:2016xwq} for the
$\rho_A$ obtained after tracing over one ghost-spin.

When the spins are entangled with the ghost-spins, then this
straightforward correlation between positive norm states and
positivity of the entanglement entropy appears to not be true as we
discuss from sec.~\ref{sec:1+2-entangled} onwards. The von
Neumann entropy contains components of $\rho_A$ which in turn contains
linear sub-combinations of the norm of the state. Thus even for
positive norm states, some components of $\rho_A$ can be negative in
general (while keeping positive the trace of $\rho_A$, which is the
norm of the state): this leads to new entanglement patterns in
general\footnote{One can restrict to subcases for which $\rho_A$ is
  diagonal (for calculational simplicity): this still leaves many
  parameters and therefore many entanglement patterns.}. Requiring 
that positive norm states give positive entanglement $S_A$ amounts to
requiring that the components $(\rho_A)^{IJ}$ are positive ($I,J$
being labels for the remaining spin variables): this is only true for
specific subregions of the Hilbert space, \ie\ only certain families
of states. (Correspondingly, negative norm states give negative real
part for $S_A$ only for certain families of states.)

More generally, when the spins are entangled with ghost-spins, we can
restrict to subfamilies of states which have correlated ghost-spins,
\ie\ the ghost-spin values are the same in each basis state. When the
number of ghost-spins is even, this implies that all allowed states
are positive norm, \ie\ negative norm states are excluded. This
restricts to half the space of states which are now all positive norm,
and the entanglement entropy is manifestly positive.  The intuition
here is in a sense akin to simulating \eg\ the $X^{\pm} + bc$
subsector of the 2-dim sigma model representing the string worldsheet
theory: in general negative norm states are cancelled between
$X^{\pm}$ and the $bc$-ghost subsectors in the eventual physical
theory. The more general subsectors in the Hilbert space where
$\rho_A$ gives positive entanglement entropy for positive norm states
can then be interpreted as the component of the state space that is
connected to this correlated ghost-spin sector. We demonstrate
this explicitly in sec.~\ref{sec:1+2-entangled} where we elaborately
study the case of a single spin entangled with two ghost-spins.

In general, the family of entangled spin \& ghost-spin systems splits
into two sectors.  One of them is with an even number of ghost-spins
and the other is with an odd number of ghost spins.  Unlike the even
ghost-spin sector, we find that for systems with odd number of
ghost-spins, such a consistent subfamily of correlated ghost-spin
states does not exist so it is not possible to uniformly pick a family
of entangled states mentioned above such that positive norm states
give positive entanglement entropy.  We analyse the case of one spin
entangled with one ghost-spin in detail in
sec.~\ref{sec:one-spin-entangled} illustrating this.  A similar
analysis for a system of $k$ spins entangled with one ghost-spin is
discussed in sec.~\ref{sec:k+1}. In sec.~\ref{sec:multi-ghost-spin},
we study systems containing multiple ghost-spins focussing on the case
of odd numbers of ghost-spins and show that there always exist
positive norm states that lead to entanglement entropy with negative
real part.

The indefinite inner products we use for the ghost-spins can be 
recast, as in the $bc$-ghost system, in terms of a ghost ``zero mode'' 
operator insertion. Then the basis states have zero norm and 
expectation values are nonvanishing only in the presence of the 
zero mode insertion (which in the $bc$-system was required to cancel 
background charge). For our present purposes in this paper, we continue 
to use the indefinite norm language.

\vspace{2mm}

\section{Reviewing spins and ghost-spins}\label{sec:revi-spins-ghost}

Here we review the toy model of two ``ghost-spins''
\cite{Narayan:2016xwq}, which abstracts away from the specific
technical issues of the ghost CFTs there but mimics some of the key
features.

Firstly, for ordinary spin variables with a 2-state Hilbert space 
consisting of\ $\{ \uparrow,\ \downarrow \}$, we take the usual positive 
definite norms in the Hilbert space\ 
\be
{\rm spins}:\qquad\qquad\qquad \langle \uparrow | \uparrow\rangle = 
\langle \downarrow | \downarrow\rangle = 1\ ,\qquad 
\langle \uparrow | \downarrow\rangle = 
\langle \downarrow | \uparrow\rangle = 0\ .
\ee
A generic state, its adjoint and positive definite norm are
\be\label{spin-psiadjnorm}
|\psi\rangle = c_1 | \uparrow\rangle + c_2 | \downarrow\rangle\ ,\qquad
{\rm adjoint}\ \ 
\langle\psi| = c_1^* \langle\uparrow| + c_2^* \langle\downarrow|\ ;\qquad 
\langle \psi| \psi\rangle = |c_1|^2 + |c_2|^2\ .
\ee
Thus we can normalize states as\ $\langle \psi| \psi\rangle = 1$
and pick a representative ray with unit norm (equivalent to calculating 
expectation values of operators as\ $\langle O \rangle 
= {\langle \psi| O |\psi\rangle \over \langle \psi| \psi\rangle} = 
\langle \psi| O |\psi\rangle$).\ 
The reduced density matrix obtained by tracing out the second spin is\
\be\label{redDM}
\rho_A = tr_B |\psi\rangle \langle \psi| = \sum_i \langle i_B|\psi\rangle 
\langle \psi| i_B\rangle = \langle \uparrow_B|\psi\rangle 
\langle \psi| \uparrow_B\rangle 
+ \langle \downarrow_B|\psi\rangle \langle \psi| \downarrow_B\rangle\ .
\ee
The familiar discussions in 2-spin systems of entanglement entropy 
via the reduced density matrix are recovered as follows.
States of the system such as\ \ $|\psi\rangle = c_1 |\uparrow \uparrow\rangle 
+ c_2 |\downarrow \downarrow\rangle$\ can be normalized as\
$\langle \psi| \psi\rangle = 1 = |c_1|^2 + |c_2|^2$ which is 
positive definite, and ensure that $|c_1|, |c_2| \leq 1$.\ With 
these norms, the reduced density matrix (\ref{redDM}) becomes
$\rho_A = |c_1|^2 | \uparrow\rangle \langle \uparrow |\ +\ 
|c_2|^2 | \downarrow\rangle \langle \downarrow |$.\ Note that 
the reduced density matrix is automatically normalized as\ 
$tr \rho_A = 1$ once the state $|\psi\rangle$ is normalized. Thus 
the entanglement entropy given as the von Neumann entropy of $\rho_A$ 
is\ $S_A = -tr \rho_A \log \rho_A = -\sum_i \rho_A(i) \log \rho_A(i)$ 
which is positive definite since each eigenvalue $\rho_A(i) < 1$ 
makes the $-\log\rho_A(i) > 0$. For the states above with $x=|c_1|^2$, 
we obtain $S_A=-x\log x-(1-x)\log(1-x)$:\ this is positive definite 
since $0<x<1$.

We define a single ``ghost-spin'' by a similar 2-state Hilbert space\ 
$\{ \uparrow,\ \downarrow \}$, but with norms 
\be\label{ghost-norms}
{\rm ghost\ spins}:\qquad\qquad\qquad \langle \uparrow | \uparrow\rangle = 
\langle \downarrow | \downarrow\rangle = 0\ ,\qquad\quad
\langle \uparrow | \downarrow\rangle = 
\langle \downarrow | \uparrow\rangle = 1\ .
\ee
This is akin to the normalizations in the $bc$-ghost system in
\cite{Narayan:2016xwq} (see \eg\ \cite{polchinskiTextBk}, Appendix, 
vol.~1 where this inner product appears). Now a generic state and 
its non-positive norm are
\be\label{<psi|psi>}
|\psi\rangle = c_1 |\uparrow\rangle + c_2 |\downarrow\rangle \qquad 
\Rightarrow\qquad \langle \psi| \psi \rangle = c_1 c_2^* + c_2 c_1^*\ ,
\ee
where we have taken the adjoint to be 
$\langle\psi| = c_1^* \langle\uparrow| + c_2^* \langle\downarrow|$, 
as in (\ref{spin-psiadjnorm}).\ Then for instance\
$|\uparrow\rangle - |\downarrow\rangle$ has norm $-2$.\ It is then 
convenient to change basis to
\be\label{ghostbasis+-}
|\pm\rangle \equiv {1\over \sqrt{2}} \big(|\uparrow\rangle\ \pm\
|\downarrow\rangle \big)\ ,
\qquad
\langle +| + \rangle = \gamma_{++} = 1 ,\quad 
\langle -| - \rangle = \gamma_{--} = -1 ,
\quad \langle +| - \rangle = \langle -| + \rangle = 0 .
\ee
A generic state with nonzero norm can be normalized to norm $+1$ or $-1$. 
Then a negative norm state can be written as\ 
$|\psi\rangle = \psi^+ |+\rangle + \psi^- |-\rangle$ with 
$\langle \psi| \psi\rangle = |\psi^+|^2 - |\psi^-|^2 = -1$.
For every such state (or ray), there is a corresponding state\ 
$|\psi^\perp\rangle$ with norm $+1$ orthogonal to $|\psi\rangle$, \ie\
$\langle \psi^\perp| \psi^\perp\rangle = 1,\ 
\langle \psi^\perp| \psi\rangle = 0$.
There are also zero norm states with $\langle \psi| \psi\rangle = 0$, 
\ie\ $|\psi^+|^2 = |\psi^-|^2$,
which do not admit any canonical normalization, \eg\ 
$|\uparrow\rangle,\ |\downarrow\rangle$.

Now considering the two ghost-spin system, basis states are
\be\label{basisstates+-}
|s_A s_B\rangle\ \equiv\ 
|\uparrow \uparrow\rangle , \ |\uparrow \downarrow\rangle ,\ 
|\downarrow \uparrow\rangle , \ |\downarrow \downarrow\rangle\ \ \ \ 
\equiv\ \ \ | + + \rangle ,\ | + - \rangle ,\ | - + \rangle ,\ 
| - - \rangle\ .
\ee
The $|\pm\pm\rangle$ basis is more transparent for our purposes. 
The inner product or metric on this space of states is not positive 
definite so the various contractions need to be defined carefully. We 
define the states, adjoints and norms as
\bea
&& |\psi\rangle = \sum \psi^{\alpha\beta} |\alpha\beta\rangle\ ,\qquad\qquad 
{\rm adjoint}:\quad\langle\psi| = \sum \langle \alpha\beta| {\psi^{\alpha\beta}}^*\ ,\nonumber\\
&& \qquad \langle\psi|\psi\rangle\ = \langle \kappa| \alpha\rangle \langle \lambda| \beta{}\rangle 
\psi^{\alpha\beta} {\psi^{\kappa\lambda}}^*\ \equiv\ 
\gamma_{\alpha\kappa} \gamma_{\beta\lambda} \psi^{\alpha\beta} {\psi^{\kappa\lambda}}^*\ 
=\ \gamma_{\alpha\alpha} \gamma_{\beta\beta} |\psi^{\alpha\beta}|^2 \ ,
\eea
where repeated indices as usual are summed over: the last expression 
pertains to the $|\pm\rangle$ basis where the metric $\gamma$ is diagonal, 
with\ $\gamma_{++}=1,\ \gamma_{--}=-1$.\
A generic normalized positive/negative norm state with norm $\pm 1$ is
\bea\label{psiijNorm1}
&& |\psi\rangle =  \psi^{++} | + + \rangle\ + \psi^{+-} | + - \rangle\ 
+ \psi^{-+} | - + \rangle\ + \psi^{--} | - - \rangle \nonumber\\ 
&& \qquad\Rightarrow\qquad 
\langle\psi| \psi\rangle = |\psi^{++}|^2 + |\psi^{--}|^2 
- |\psi^{+-}|^2 - |\psi^{-+}|^2 = \pm 1\ .
\eea
This translates to corresponding conditions on the coefficients
$\psi^{\alpha\beta}$.
A simple example of a positive norm state is\
$|\psi\rangle = \psi^{++} | + + \rangle\ + \psi^{--} | - - \rangle$,\
while\ $|\psi\rangle =  \psi^{+-} | + - \rangle\ + \psi^{-+} | - + \rangle$\ 
has negative norm.

With the density matrix $\rho=|\psi\rangle\langle\psi| 
= \sum \psi^{\alpha\beta} {\psi^{\kappa\lambda}}^* | \alpha\beta \rangle\langle \kappa\lambda|$, the reduced 
density matrix obtained by a partial trace over one spin can again be 
defined via a partial contraction as
\be
\rho_A = tr_B \rho \equiv (\rho_A)^{\alpha\kappa} |\alpha\rangle \langle \kappa|\ ,\qquad\ \  
(\rho_A)^{\alpha\kappa}\ =\ \gamma_{\beta\lambda} \psi^{\alpha\beta} {\psi^{\kappa\lambda}}^*\ =\ 
\gamma_{\beta\beta}  \psi^{\alpha\beta} {\psi^{\kappa\beta}}^*\ ,
\ee
\bea\label{rhoAc1234}
\Rightarrow\qquad\qquad 
(\rho_A)^{++} =\ |\psi^{++}|^2 - |\psi^{+-}|^2\ , &\quad&
(\rho_A)^{+-} =\ \psi^{++} {\psi^{-+}}^* - \psi^{+-} {\psi^{--}}^*\ , 
\nonumber\\
(\rho_A)^{-+} =\ \psi^{-+} {\psi^{++}}^* - \psi^{--} {\psi^{+-}}^*\ , &\quad& 
(\rho_A)^{--} =\ |\psi^{-+}|^2 - |\psi^{--}|^2\ , 
\eea
Then $tr \rho_A = \gamma_{\alpha\kappa} (\rho_A)^{\alpha\kappa} = (\rho_A)^{++} - (\rho_A)^{--}$. 
Thus the reduced density matrix is normalized to have\ 
$tr \rho_A = tr \rho = \pm 1$ depending on whether the state 
(\ref{psiijNorm1}) is positive or negative norm.
Also, $\rho_A$ can have some eigenvalues negative.

The entanglement entropy calculated as the von Neumann entropy of 
$\rho_A$ is
\be\label{EErhoA}
S_A = -\gamma_{\alpha\beta} (\rho_A \log \rho_A)^{\alpha\beta}\ =\ 
- \gamma_{++} (\rho_A \log \rho_A)^{++}
- \gamma_{--} (\rho_A \log \rho_A)^{--}
\ee
where the last expression pertains to the $|\pm\rangle$ basis with 
$\gamma_{\pm\pm}=\pm 1$.  This requires defining $\log\rho_A$ as an 
operator: we define this as the usual $log$-expansion
\be
(\log\rho_A)^{\alpha\kappa} = (\log(1+\rho_A-1))^{\alpha\kappa} 
= (\rho_A-1)^{\alpha\kappa} - {1\over 2}(\rho_A-1)^{\alpha\beta}
\gamma_{\beta\lambda}(\rho_A-1)^{k} + \ldots
\ee
or equivalently via\ $(\rho_A)^{\alpha\kappa} = (e^{\log\rho_A})^{\alpha\kappa}
= 1^{\alpha\kappa} + (\log\rho_A)^{\alpha\kappa} + {1\over 2!}
(\log\rho_A)^{\alpha\beta}\gamma_{\beta\lambda}(\log\rho_A)^{\lambda\kappa} 
+ \ldots$\ and the solution thereof.\ The signs in the contractions 
in $\log\rho_A$ are perhaps more easily dealt with if we use the 
mixed-index reduced density matrix $(\rho_A)^\alpha{_\kappa}$.

To illustrate this, let us for simplicity consider a simple family of 
states where the reduced density matrix is diagonal, by restricting to 
${\psi^{-+}}^* = {\psi^{+-} {\psi^{--}}^*\over \psi^{++}}$. In this case, 
$\log\rho_A$ is also diagonal and can be calculated easily. 
From (\ref{rhoAc1234}) for the state (\ref{psiijNorm1}), this gives
\bea
{\psi^{-+}}^* = {\psi^{+-} {\psi^{--}}^*\over \psi^{++}}\ \ \ &&
\Rightarrow\qquad
\langle\psi| \psi\rangle = (|\psi^{++}|^2 - |\psi^{+-}|^2) 
\Big(1 + {|\psi^{--}|^2\over |\psi^{++}|^2} \Big) = \pm 1\ , \nonumber\\
\rho_A &=& \pm \left[ {|\psi^{++}|^2\over |\psi^{++}|^2 + |\psi^{--}|^2} 
|+ \rangle\langle +|\
-\ {|\psi^{--}|^2\over |\psi^{++}|^2 + |\psi^{--}|^2} 
|- \rangle\langle -| \right] ,\qquad\ \ 
\eea
where the $\pm$ pertain to positive and negative norm states respectively.
The location of the negative eigenvalue is different for positive and 
negative norm states, leading to different results for the von Neumann 
entropy. For negative norm states, $(\rho_A)^{++} < 0,\ (\rho_A)^{--} > 0$.\
Then the mixed-index reduced density matrix components\ 
$(\rho_A)_\alpha^\kappa = \gamma_{\alpha\beta} (\rho_A)^{\beta\kappa}$\ are
\be
(\rho_A)^+_+ = \pm x\ ,\qquad (\rho_A)^-_- = \pm (1-x)\ ,\qquad\ 
x = {|\psi^{++}|^2\over |\psi^{++}|^2 + |\psi^{--}|^2}\ ,\qquad 0 < x < 1\ .
\ee
Thus\ $tr\rho_A = (\rho_A)^+_+ + (\rho_A)^-_- = \pm 1$ 
manifestly. Now we obtain\ 
$(\log\rho_A)^+_+ = \log (\pm x)$ and $(\log\rho_A)^-_- = \log (\pm (1-x))$,
the $\pm$ referring again to positive/negative norm states respectively. 
The entanglement entropy (\ref{EErhoA}) becomes\
$S_A = - (\rho_A)^+_+ (\log\rho_A)^+_+ - (\rho_A)^-_- (\log\rho_A)^-_-$\ 
and so
\bea\label{2gsEE}
\langle\psi| \psi\rangle > 0: &&  S_A = - x\log x - (1-x) \log (1-x) > 0\ ,\\
\langle\psi| \psi\rangle < 0: && S_A =  x\log (-x) + (1-x) \log (-(1-x)) 
= x\log x + (1-x) \log (1-x) \nonumber \\
&&\phantom{S_A}= x\log x + (1-x) \log (1-x) + i\pi(2n+1)x + i\pi(2m+1)(1-x)\ .
\nonumber  
\eea
For positive norm states, $S_A$ is manifestly positive since $x<1$, 
just as in an ordinary 2-spin system.\ For negative norm states, 
we note that for the principal branch, \ie\ $n=m$, the imaginary part 
is independent of $x$, \ie\ the same for all such negative norm states 
if we choose the same branch for the logarithms. In our analysis that 
follows, we will for simplicity consider the principal branch only (with 
$n,m=0$), \ie\ we will effectively set $\log (-1)=i\pi$ henceforth.
The real part of entanglement entropy is negative since $x<1$ and 
the logarithms are negative: apart from the minus sign, it is the same 
as $S_A$ for the positive norm states. This real part is minimized 
when $x={1\over 2}$\ (this value corresponds to maximal entanglement 
for positive norm states): this ``minimal'' entanglement is 
$S_A = -\log 2 + i\pi$.

The above discussion can also be phrased in terms of the
$|\uparrow\rangle, |\downarrow\rangle$ basis although we have found it
convenient to use the $|\pm\rangle$ basis. It is worth noting that
while (\ref{ghost-norms}) mimics the norms of the $bc$-ghost system in
\cite{Narayan:2016xwq}, there is no obvious analog of the background
charge here: in particular tracing over spin$_A$ instead of spin$_B$
is equivalent, so that entanglement entropy for the subsystem is the
same as that for the complement.

\section{Tracing over ghost-spins:  the reduced density matrix}
\label{sec:rdm-gs}

We consider systems of spins and ghost-spins, possibly entangled. The
ghost-spins, representing the negative norm states, are regarded as
invisible. The physical system is represented by the spin degrees of 
freedom and the physical information content thereof is obtained by 
tracing over the ghost-spins.

Operationally, we start with a state $|\psi\rangle$ in the full
Hilbert space, and the corresponding full density matrix $\rho =
|\psi\rangle\langle\psi|$ and construct a reduced density matrix by
tracing over all the ghost-spins, \ie\ $\rho^s = {\rm tr}_{gs}
(\rho)$. The resulting subsystem is now a mixed state described by the
reduced density matrix $\rho^s$. Since this comprises only physical
spin variables, we must require that this be a well-defined physically
sensible system. As a minimal requirement in this regard, we expect
that any observable $O_s$ of the spin variables alone has a 
correlation function in the state $|\psi\rangle$ that must satisfy
\be\label{rdm-gs}
\langle\psi| O_s |\psi\rangle = tr_s (O_s \rho^s)\ .
\ee
Here the left hand side is the correlation function calculated in the 
full state $|\psi\rangle$ (including the ghost-spins), while the right 
hand side is calculated in the mixed state $\rho^s$ describing the 
remaining spins obtained after tracing over the ghost-spins. Since 
the left hand side contains an implicit trace over the ghost-spins, 
this gives a definition for the reduced density matrix $\rho^s$. When 
$O_s$ is the identity operator, (\ref{rdm-gs}) fixes the normalization 
of $\rho^s$ as
\be\label{rdm-gsIdOp}
\langle\psi|\psi\rangle = tr_s (\rho^s)\ .
\ee
In particular for positive norm states normalized as 
$\langle\psi|\psi\rangle = 1$, we have $tr_s (\rho^s) = 1$. The 
correlation function above of course appears in the expectation value 
of the observable as $\langle O_s\rangle = 
{\langle\psi| O_s |\psi\rangle\over \langle\psi|\psi\rangle}$ and 
so it differs from the expectation value by the sign of the norm 
$\langle\psi|\psi\rangle$ of the state (which is negative for negative 
norm states)\footnote{To see that these expressions are consistent, 
consider a simple example of spins disentangled from ghost-spins 
(which we discuss in more detail in sec.~\ref{sec:spins-disent-ghost}). 
The state $|\psi\rangle$ can then be written as a product state 
$|\psi\rangle=|\psi_s\rangle|\psi_{gs}\rangle$ and its norm is 
$\langle\psi|\psi\rangle = \langle\psi_s|\psi_s\rangle 
\langle\psi_{gs}|\psi_{gs}\rangle$. We normalize the norm as
$\langle\psi|\psi\rangle = \pm 1$ for positive/negative norm states 
respectively. The expectation 
value is $\langle O_s\rangle = {\langle\psi| O_s |\psi\rangle\over
\langle\psi|\psi\rangle} = {\langle\psi_s| O_s |\psi_s\rangle 
\langle\psi_{gs}|\psi_{gs}\rangle\over
\langle\psi_s|\psi_s\rangle\langle\psi_{gs}|\psi_{gs}\rangle } = 
{\langle\psi_s| O_s |\psi_s\rangle\over\langle\psi_s|\psi_s\rangle}$~. 
Then $tr_s(O_s\rho^s) = \langle\psi| O_s |\psi\rangle = 
\langle\psi_s| O_s |\psi_s\rangle \langle\psi_{gs}|\psi_{gs}\rangle = 
\pm {\langle\psi_s| O_s |\psi_s\rangle \over \langle\psi_s|\psi_s\rangle}
= \pm \langle O_s\rangle$. In particular for $O_s$ the identity 
operator, we have $tr_s\rho^s = \pm 1 = tr\rho = \langle\psi|\psi\rangle$ 
as expected.}.

To illustrate this, consider a simple example of one spin and two 
ghost-spins, in a state 
\be
|\psi\rangle = \psi^{i,\al\beta} |i\rangle |\al\beta\rangle\ ,
\ee
where $i$ labels the spin variable and $\al,\beta$ the ghost-spins. 
Consider an observable $O_s$ of the spin variable alone: this can be
written as $O_s=O_s^{ij} |i\rangle\langle j|$. Its correlation function 
is
\bea
&& \langle\psi| O_s |\psi\rangle = (\psi^*)^{j,\sigma\rho} 
\langle \sigma\rho|\langle j| O_s^{kl} |k\rangle\langle l| 
\psi^{i,\al\beta} |i\rangle |\al\beta\rangle 
= (\psi^*)^{j,\sigma\rho} O_s^{kl} \psi^{i,\al\beta}\ \langle j |k\rangle\
\langle l |i\rangle\ \langle \sigma\rho |\al\beta\rangle\quad \nonumber\\
&& \qquad\qquad  \equiv g_{jk} g_{li}\ O_s^{kl} (\rho^s)^{ij} 
\equiv tr (O_s \rho^s)
\eea
where $g_{jk} = \langle j |k\rangle$ is the positive definite inner 
product for the spin states. The above expression has traced over the 
ghost-spins and shows the resulting reduced density matrix to be
\be\label{rdm-gs-rho}
(\rho^s)^{ij} = \gamma_{\al\sigma} \gamma_{\beta\rho} 
\psi^{i,\al\beta} (\psi^*)^{j,\sigma\rho} = 
\gamma_{\al\al} \gamma_{\beta\beta} \psi^{i,\al\beta} (\psi^*)^{j,\al\beta}\ ,
\ee
where $\langle \sigma|\al \rangle = \gamma_{\al\sigma} = \gamma_{\sigma\al}$ 
is the indefinite inner product over the ghost-spin states. In our 
analysis (and as in \cite{Narayan:2016xwq}), we assume that the 
ghost-spin states have an inner product given by a real-valued, 
symmetric metric $\gamma_{\al\beta}$. In particular, as reviewed earlier, 
in the $|\pm\rangle$ basis (\ref{ghostbasis+-}), we have\ 
$\gamma_{++}=1,\ \gamma_{--}=-1$. The second expression in (\ref{rdm-gs-rho}) 
is specific to this diagonal metric.
The resulting reduced density matrix still needs to satisfy positivity 
properties, if it is to describe a physical spin system: this 
imposes various conditions on generic states comprising entangled 
spins and ghost-spins, as we will discuss at length in what follows.

For more general spin and ghost-spin systems, the above discussion 
can be generalized as follows. A generic state is 
$|\psi\rangle = \psi^{I,\al} |I\rangle|\al\rangle$, where $I$ is a 
collective label for states of multiple spin variables, and $\al$ is a
collective label for states of multiple ghost-spins. Then 
\be\label{rdm-gs-rhoGen}
\begin{split}
\langle\psi| O_s |\psi\rangle 
= (\psi^*)^{J,\sigma} O_s^{KL} \psi^{I,\al}\ \langle J |K\rangle\
\langle L |I\rangle\ \langle \sigma |\al\rangle 
\equiv tr (O_s \rho^s)\\  \Rightarrow\qquad 
(\rho^s)^{IJ} = \langle \sigma |\al\rangle \psi^{I,\al} (\psi^*)^{J,\sigma}\ ,
\qquad
\end{split}
\ee
where $\langle J |K\rangle$ symbolises a product of multiple 
individual inner products $\prod \langle j |k\rangle$ and 
$\langle \sigma |\al\rangle$ a product of multiple ghost-spin inner 
products $\prod \langle \sigma_k|\al_k \rangle$. This defines the reduced 
density matrix as above,
 which now has multiple indices $I,J$.

As we see, the contractions in the reduced density matrix are fixed as
given above, and roughly speaking they are correlated with the
contraction patterns in the norm of the state. One might ask if there
are other contraction schemes that one may cook up formally to trace
over the ghost-spins towards defining the reduced density matrix. For
instance consider
\be\label{rdm12-2}
(\rho_A)^{ik} = \gamma_{\al\rho} \gamma_{\beta\sigma} 
\psi^{i,\al\beta} (\psi^*)^{k,\sigma\rho} = 
\gamma_{\al\al} \gamma_{\beta\beta} \psi^{i,\al\beta} (\psi^*)^{k,\beta\al}\ ,
\ee
in the case of one spin and two ghost-spins. Here the complex 
conjugated state has reversed index list, instead of (\ref{rdm-gs-rho}) 
where the conjugated state has first index contracted with the metric 
$\gamma_{\al\beta}$. While this appears consistent formally, it does not 
satisfy the physical conditions (\ref{rdm-gs}) and (\ref{rdm-gsIdOp}),
that must hold for the residual physical subsystem of spin variables
alone. We will comment on this in specific places in what follows.

To summarise, we have seen how the reduced density matrix
(\ref{rdm-gs}), (\ref{rdm-gs-rho}), (\ref{rdm-gs-rhoGen}), arises from
tracing over the (invisible) ghost-spins, satisfying the physical
requirements (\ref{rdm-gs}), (\ref{rdm-gsIdOp}) expected of the
residual physical spin system. In what follows, we will explore the
patterns of entanglement that arise from this operation in various
categories of spin \& ghost-spin systems.

\section{Spins disentangled from ghost-spins}\label{sec:spins-disent-ghost}

We will start with a configuration where spins are not entangled with
ghost spins.  This case is similar in spirit to the longitudinal and
time-like degrees of freedom and their ghost counterparts in gauge
theories.  In the free theory, these sectors are decoupled from the
physical sector.  As we will see, in the disentangled spin/ghost-spin
system we can show in general that the entanglement entropy is
positive definite for positive norm states and has negative definite
real part for negative norm states.

We define the norm of a state in the Hilbert space by defining the
metric on the state space.  The spin Hilbert space has a positive
definite metric while the ghost-spin states have a non-positive metric
$\gamma_{\alpha\beta}$,
\be
g_{ij}=\delta_{ij}\ ,\qquad\qquad \gamma_{++}=1,\ \ \gamma_{--}=-1\ . 
\ee
The $\gamma_{\alpha\beta}$ metric is the same as (\ref{ghostbasis+-}), 
equivalent to the off-diagonal form\ 
$\langle\uparrow |\downarrow\rangle = 
\langle\downarrow |\uparrow\rangle = 1$\ in (\ref{ghost-norms}).\
This is equivalent to defining the adjoints of the ghost-spin states as
\be
\left( | \uparrow\rangle \right)^\dag = \langle \uparrow| 
= \langle \downarrow| c_0\ , \qquad\qquad 
\left( | \downarrow\rangle \right)^\dag = \langle \downarrow| 
= \langle \uparrow| c_0\ ,
\ee
with $c_0$ a ``zero mode'' operator, analogous to the ghost zero mode
$c_0$ in the $c=-2$ $bc$-ghost system \cite{Narayan:2016xwq}, where 
nonvanishing correlation functions required an appropriate ghost zero
mode insertion to cancel the background charge inherent in these
systems \footnote{In the present context also, this implies the 
existence of a pair of operators satisfying\ $\{b_0, c_0\}=1$ with 
the $\{ \uparrow, \downarrow\}$ states forming a representation thereof.
We then couple this with the usual positive definite norm on states in
the Hilbert space to define expectation values. Nonzero expectation 
values are obtained only after a $c_0$ insertion: \ie\
$\langle\downarrow |\downarrow\rangle = 0 = \langle\uparrow |\uparrow\rangle ,
\ \langle \downarrow |c_0 | \downarrow\rangle = 1 = 
\langle\uparrow|c_0|\uparrow\rangle$.\  A generic ghost-spin state\ 
$|\psi\rangle = c_1 |\uparrow\rangle + c_2 |\downarrow\rangle$\ 
then has adjoint\
$\left( |\psi\rangle \right)^\dag = c_1^* \langle \downarrow|c_0
+ c_2^* \langle \uparrow|c_0$,\ 
recovering the inner product\
$\left( (|\psi\rangle)^\dag, |\psi\rangle \right) \equiv 
\langle \psi| \psi \rangle = c_1 c_2^* + c_2 c_1^*$ identical to 
(\ref{<psi|psi>}).
Thus our analysis can equivalently be phrased using the explicit 
insertion of this $c_0$ operator in expectation values: in this 
rephrasing, all expectation values vanish without the insertion and 
entanglement entropy also vanishes.
An analog of the ghost-number operator here would be $N_g\sim c_0b_0$, 
which can be used to classify states. Alongwith this, an analog of 
the Hamiltonian $L_0$ with appropriate commutation relations would 
be useful to study dynamics in these systems: our study here is mostly 
``kinematic''.}. In our discussion throughout this paper, we will 
however continue to use the non-positive metric for ghost-spin states 
for simplicity.

Returning to our discussion of entanglement entropy, if the spin sector 
is not entangled with the ghost-spin sector, then the most general state 
is of product form
\bea
|\psi\rangle = |\psi_s\rangle\ |\psi_{gs}\rangle\ ,\qquad && \quad
\langle\psi|\psi\rangle = \langle\psi_s|\psi_s\rangle\ 
\langle\psi_{gs}|\psi_{gs}\rangle\ ,\\
\langle\psi_s|\psi_s\rangle = g_{i_1j_1} \ldots g_{i_nj_n} 
(\psi_s)^{i_1i_2\ldots} (\psi_s)^{j_1j_2\ldots *}\ ,&& 
\langle\psi_{gs}|\psi_{gs}\rangle = \gamma_{\alpha_1\beta_1} \ldots \gamma_{\alpha_n\beta_n} 
(\psi_{gs})^{\alpha_1\alpha_2\ldots} (\psi_{gs})^{\beta_1\beta_2\ldots *} .\ \nonumber
\eea
Since $|\psi_s\rangle$ is contracted with $g_{ij}$, this sector is 
entirely positive norm as expected (with 
$\langle\psi_s|\psi_s\rangle > 0$): on the other hand, $|\psi_{gs}\rangle$ 
contracted with $\gamma_{\alpha\beta}$ can give rise to negative norm states if 
$\langle\psi_{gs}|\psi_{gs}\rangle < 0$. 

The reduced density matrix obtained after tracing over all the ghost-spins 
is
\bea\label{rhoAsgsDisent}
&& \qquad\qquad
\rho_A^s = tr_{gs} \big( |\psi_s\rangle\ |\psi_{gs}\rangle \langle\psi_s|\ 
\langle\psi_{gs}| \big)\ ,\\
&&\!\!\!\! (\rho_A^s)^{i_1\ldots , k_1\ldots} = \gamma_{\alpha_1\beta_1}\ldots \gamma_{\alpha_n\beta_n}\
(\psi_s)^{i_1\ldots} (\psi_s)^{k_1\ldots *}\ 
(\psi_{gs})^{\alpha_1\ldots} (\psi_{gs})^{\beta_1\ldots *} = 
\langle\psi_{gs}|\psi_{gs}\rangle\ 
(\psi_s)^{i_1\ldots} (\psi_s)^{k_1\ldots *} .\ \nonumber
\eea
We will now normalize positive/negative norm states to have norm $\pm 1$ 
respectively, \ie\
\be
\langle\psi_{gs}|\psi_{gs}\rangle \gtrless 0\qquad\Rightarrow\qquad
\langle\psi|\psi\rangle = \langle\psi_s|\psi_s\rangle\ 
\langle\psi_{gs}|\psi_{gs}\rangle = \pm 1\qquad\qquad 
[\langle\psi_s|\psi_s\rangle > 0]\ .
\ee
From (\ref{rhoAsgsDisent}), we then see that $\rho_A^s$ is automatically 
normalized as
\be
(\rho_A^s)^{i_1\ldots , k_1\ldots} = \pm {1\over \langle\psi_s|\psi_s\rangle} 
(\psi_s)^{i_1\ldots} (\psi_s)^{k_1\ldots *} \qquad\Rightarrow\qquad
tr \rho_A^s = \pm 1 \qquad \big(\langle\psi|\psi\rangle \gtrless 0\big)\ .
\ee
For positive norm states, $\rho_A^s$ is positive definite with
eigenvalues $0<\lambda_i< 1$\ (if the spin sector is entangled) and\
$\sum_i\lambda_i=1$: thus we have the usual positive definite
entanglement entropy for $\rho_A^s$,
\be
S_A = - {\rm tr}_s\ \rho_A^s \log\rho_A^s 
= -\sum_i \lambda_i\log\lambda_i > 0\ .
\ee
For negative norm states on the other hand, $\rho_A^s$ is negative 
definite, with eigenvalues $-\lambda_i$.\ This gives
\be\label{-venormDisentEE}
S_A = - {\rm tr}_s\ \rho_A^s \log\rho_A^s 
= -\sum_i (-\lambda_i)\log(-\lambda_i) = \sum_i \lambda_i\log\lambda_i 
+ i\pi\ ,
\ee
with a negative definite real part and constant imaginary part (since
$\sum_i\lambda_i=1$). As mentioned after (\ref{2gsEE}), this constant
imaginary part here stems from our choice of the branch of the
logarithm with $\log(-1)=i\pi$ (for simplicity) and corresponds to an
overall minus sign in the reduced density matrix (which is otherwise
positive definite, with no relative minus sign amongst the eigenvalues).

Thus when the spin sector is not entangled with the ghost-spin sector
(both of which can be entangled within themselves), we see in great
generality that positive norm states have positive entanglement entropy 
while negative norm states have entanglement entropy with a negative 
definite real part and a constant imaginary part. We recall that the 
two ghost-spin system exhibited similar behaviour \cite{Narayan:2016xwq}.

\subsection{Two spins and two ghost-spins, disentangled}

We will illustrate the above generalities for a simple but illustrative 
example: consider a system of two spins and two ghost spins.
The general state in this case and its norm are
\be
|\psi\rangle = \psi^{ij,\al\beta} |ij\rangle |\al\beta\rangle\ ,\qquad
\langle\psi|\psi\rangle = g_{ik} g_{jl} \gamma_{\al\sigma} \gamma_{\beta\rho} 
\psi^{ij,\al\beta}(\psi^*)^{kl,\sigma\rho}\ .
\ee
Since there are two $\gamma_{\alpha\beta}$ factors in the contraction, it is 
clear that terms with a single minus ghost-index $\al,\beta$ will 
acquire a minus sign, giving \eg\ $\big(-|\psi^{+-,+-}|^2\big)$ while 
terms like $\big(|\psi^{++,--}|^2\big)$ will contribute with a $+$-sign 
in the norm.

From sec.~\ref{sec:rdm-gs}, tracing over the ghost-spins gives the 
reduced density matrix ((\ref{rdm-gs}), (\ref{rdm-gs-rho}), 
(\ref{rdm-gs-rhoGen}))
\be\label{rhoA22-1}
(\rho_A)^{ij,kl} = \gamma_{\al\sigma} \gamma_{\beta\rho} 
\psi^{ij,\al\beta} (\psi^*)^{kl,\sigma\rho} = 
\gamma_{\al\al} \gamma_{\beta\beta} \psi^{ij,\al\beta}
(\psi^*)^{kl,\al\beta}\ .
\ee

Taking the spins to be disentangled from the 
ghost-spins (both of which could be entangled within themselves), 
the general state is of product form,
\bea\label{psi22genDisent}
&& \qquad |\psi\rangle = |\psi_s\rangle |\psi_{gs}\rangle 
= \big(c_{++} |++\rangle + c_{+-} |+-\rangle + 
c_{-+} |-+\rangle + c_{--} |--\rangle\big)\times\nonumber\\ 
&&\qquad\qquad\qquad\qquad\qquad\qquad \big(
\psi^{++} |++\rangle + \psi^{+-} |+-\rangle + \psi^{-+} |-+\rangle
+ \psi^{--} |--\rangle \big) \ ,\nonumber\\
&&\!\!\! \langle\psi|\psi\rangle = \big(|c_{++}|^2 + |c_{+-}|^2 +
|c_{-+}|^2 + |c_{--}|^2\big) 
\big(|\psi^{++}|^2 - |\psi^{+-}|^2 - |\psi^{-+}|^2 +
|\psi^{--}|^2\big)\ . \quad \ \ \ \ 
\eea
Thus we have\ $\psi^{++,++}\equiv c_{++}\psi^{++}$ etc. 
Using (\ref{rhoA22-1}) gives\ 
\be\label{rhoA22genDisent}
(\rho_A)^{ij,kl} = \langle\psi_{gs}|\psi_{gs}\rangle\ c^{ij} c^{kl*}\qquad
\longrightarrow\qquad (\rho_A)^{i,k} = g_{jl} (\rho_A)^{ij,kl}\ ,
\ee
where we have performed a further trace over one of the spins to 
obtain the reduced density matrix $(\rho_A)^{i,k}$ for the remaining 
spin: this gives
\bea
(\rho_A)^{+,+} = \big(|c^{++}|^2 + |c^{+-}|^2\big) 
\langle\psi_{gs}|\psi_{gs}\rangle , \qquad
(\rho_A)^{+-} = \big(c^{++}(c^*)^{-+} + c^{+-}(c^*)^{--}\big) 
\langle\psi_{gs}|\psi_{gs}\rangle , &&  \nonumber\\
(\rho_A)^{-+} = \big(c^{-+}(c^*)^{++} + c^{--}(c^*)^{+-}\big) 
\langle\psi_{gs}|\psi_{gs}\rangle , \quad
(\rho_A)^{--} = \big(|c^{-+}|^2 + |c^{--}|^2\big) 
\langle\psi_{gs}|\psi_{gs}\rangle .\ \ \ \ &&
\eea
We see that $(\rho_A)$ inherits the sign from the ghost-spin sector. 
A simple entangled spin state, its normalization and associated 
reduced density matrix are\ 
\bea
&& |\psi_s\rangle = c^{++}|++\rangle + c^{--} |--\rangle\ ,\qquad
\langle\psi_{gs}|\psi_{gs}\rangle = 
\pm {1\over \langle\psi_s|\psi_s\rangle} 
= \pm {1\over |c^{++}|^2 + |c^{--}|^2}\ ,\nonumber\\
&& (\rho_A)^{+,+} = \pm {|c^{++}|^2\over |c^{++}|^2 + |c^{--}|^2} 
\equiv \pm x\ ,  \quad
(\rho_A)^{-,-} = \pm {|c^{--}|^2\over |c^{++}|^2 + |c^{--}|^2} 
= \pm (1-x) .\quad\ \ \ \
\eea
We have $0<x<1$. Then the entanglement entropy for this state is
\be
S_A = - (\pm x) \log (\pm x) - (\pm (1-x)) \log (\pm(1-x))
\ee
which is clearly positive definite for positive norm states (+ sign).
For negative norm states, we have $S_A=x\log x + (1-x)\log(1-x) + i\pi$, 
with a negative real part and a constant imaginary part. This verifies
the general structure stated earlier.

In the following sections we will study systems of spins entangled with
ghost-spins: this is somewhat more intricate and there are many new
entanglement patterns depending on detailed properties of the
entangled state.

\section{One spin entangled with two ghost-spins}\label{sec:1+2-entangled}

We will now consider a single spin entangled with two ghost-spins.
This system, as we will see, is quite rich in generating a spectrum of
entanglement with complex entanglement entropy with non-constant
imaginary part as well as real part correlated with the norm of the
state, \ie\ there exist families of entangled states with positive
entanglement entropy when the norm is positive.

A generic state and its norm are
\be
|\psi\rangle = \psi^{i,\al\beta} |i\rangle |\al\beta\rangle\ ,\qquad
\langle\psi|\psi\rangle = g_{ij} \gamma_{\al\sigma} \gamma_{\beta\rho} 
\psi^{i,\al\beta}(\psi^*)^{j,\sigma\rho}\ .
\ee
Since there are two $\gamma_{\alpha\beta}$ factors in the contraction,
it is  clear that terms with a single minus ghost-index $\al,\beta$ will 
acquire a minus sign, giving \eg\ $\big(-|\psi^{+,+-}|^2\big)$ while 
terms like $\big(|\psi^{+,--}|^2\big)$ will contribute with a $+$-sign 
in the norm. For instance, a simple entangled state with positive 
norm is\ $|\psi\rangle = \psi^{+,++} |+\rangle |++\rangle 
+ \psi^{-,--} |-\rangle |--\rangle$ with norm\ 
$\langle\psi|\psi\rangle = |\psi^{+,++}|^2 + |\psi^{-,--}|^2$.
Explicitly writing the most general state, we have
\bea\label{psi12mostgen}
|\psi\rangle &=& \psi^{+,++} |+\rangle |++\rangle 
+ \psi^{+,+-} |+\rangle |+-\rangle + \psi^{+,-+} |+\rangle |-+\rangle
+ \psi^{+,--} |+\rangle |--\rangle \nonumber\\
&& +\ \psi^{-,++} |-\rangle |++\rangle + \psi^{-,+-} |-\rangle |+-\rangle 
+ \psi^{-,-+} |-\rangle |-+\rangle + \psi^{-,--} |-\rangle |--\rangle\ \ \ \
\eea
with norm
\be\label{psi12mostgenNorm}
\begin{split}
  \langle\psi|\psi\rangle = |\psi^{+,++}|^2 - |\psi^{+,+-}|^2
- |\psi^{+,-+}|^2 + |\psi^{+,--}|^2\qquad \\  \ +\ |\psi^{-,++}|^2 -
|\psi^{-,+-}|^2 - |\psi^{-,-+}|^2 + |\psi^{-,--}|^2\ .
\end{split}
\ee

\noindent \emph{Patterns of entanglement:}\ \ 
As discussed in sec.~\ref{sec:rdm-gs}, tracing over the ghost-spins 
gives the reduced density matrix (\ref{rdm-gs}), (\ref{rdm-gs-rho}), 
(\ref{rdm-gs-rhoGen}),
\be\label{rhoA12-1}
(\rho_A)^{ik} = \gamma_{\al\sigma} \gamma_{\beta\rho} 
\psi^{i,\al\beta} (\psi^*)^{k,\sigma\rho} = 
\gamma_{\al\al} \gamma_{\beta\beta} \psi^{i,\al\beta} (\psi^*)^{k,\al\beta}\ .
\ee
Explicitly, this reduced density matrix after tracing over both 
ghost-spins is
\bea\label{rhoA12mostgen}
 (\rho_A)^{++} &=& |\psi^{+,++}|^2 - |\psi^{+,+-}|^2 - |\psi^{+,-+}|^2
+ |\psi^{+,--}|^2\ , \qquad \nonumber\\
(\rho_A)^{+-} &=& \psi^{+,++}(\psi^*)^{-,++} - \psi^{+,+-}(\psi^*)^{-,+-}
- \psi^{+,-+}(\psi^*)^{-,-+} + \psi^{+,--}(\psi^*)^{-,--}\ ,\nonumber\\
(\rho_A)^{-+} &=& \psi^{-,++}(\psi^*)^{+,++} - \psi^{-,+-}(\psi^*)^{+,+-}
- \psi^{-,-+}(\psi^*)^{+,-+} + \psi^{-,--}(\psi^*)^{+,--}\ , \qquad \\
(\rho_A)^{--} &=& |\psi^{-,++}|^2 - |\psi^{-,+-}|^2 
- |\psi^{-,-+}|^2 + |\psi^{-,--}|^2\ .\nonumber
\eea
\noindent \emph{Physical requirement:}\ \ After tracing over the
ghost spins, we obtain a reduced density matrix for just ordinary
spins alone. On physical grounds, this should be required to be 
positive definite for positive norm states, since these can equivalently 
be decomposed into purely physical effective positive norm basis 
states (even if there were underlying ghost-like states in the full 
system). Equivalently, since the remaining spins are ordinary spins, 
they should allow good physical interpretation for positive norm 
states with positive entanglement. (The negative norm states sector 
need not allow as clear a physical interpretation.)

Firstly, as in sec.1, if we consider the spins to be disentangled from 
the ghost-spins (both of which could be entangled within themselves), 
then the general state (\ref{psi12mostgen}) is of the form
\bea\label{psi12genDisent}
&& \quad |\psi\rangle = \big(c^+ |+\rangle + c^- |-\rangle\big)\ \big(
\psi^{++} |++\rangle + \psi^{+-} |+-\rangle + \psi^{-+} |-+\rangle
+ \psi^{--} |--\rangle \big) \ ,\nonumber\\
&& \langle\psi|\psi\rangle = \big(|c_+|^2 + |c_-|^2\big) 
\big(|\psi^{++}|^2 - |\psi^{+-}|^2 - |\psi^{-+}|^2 + |\psi^{--}|^2\big) 
 = \langle\psi_s|\psi_s\rangle\ \langle\psi_{gs}|\psi_{gs}\rangle\
 .\quad\ \ \ \ 
\eea
In other words, here\ $\psi^{+,++}\equiv c^+\psi^{++}$ etc. This gives
\bea\label{rhoA12genDisent}
&& (\rho_A)^{++} = |c^+|^2 \langle\psi_{gs}|\psi_{gs}\rangle\ , 
\qquad (\rho_A)^{+-} = c^+(c^*)^- \langle\psi_{gs}|\psi_{gs}\rangle\ ,
\nonumber\\
&& \qquad 
(\rho_A)^{-+} = c^-(c^*)^+ \langle\psi_{gs}|\psi_{gs}\rangle\ , \qquad 
(\rho_A)_{--} = |c_-|^2 \langle\psi_{gs}|\psi_{gs}\rangle\ .\nonumber
\eea
We see that $(\rho_A)$ acquires the sign from the ghost-spin sector. 
In this case, since there is just a single spin, $\det (\rho_A)=0$ and 
$S_A=0$ of course. In the two spins disentangled from two ghost-spins, 
we saw earlier that $S_A>0$ for positive norm states.

Next, we consider the cases where the spin is entangled with the two
ghost-spins.  We see from (\ref{rhoA12-1}), (\ref{rhoA12mostgen}), that
in general $\rho_A$ positivity (and thereby $S_A>0$) for positive norm
states is not possible in the entire Hilbert space of states but is
possible in subsectors thereof (\ie\ for subfamilies of states), since
there are sufficiently many parameters\footnote{For a single spin 
entangled with a ghost-spin, we will see in the next section that 
$\rho_A$ always has a negative eigenvalue: 
so this sector cannot be salvaged since there are not enough
parameters. Interestingly, for the system of two ghost-spins 
\cite{Narayan:2016xwq} reviewed earlier, this is possible: the signs 
in $\rho_A$ are just right!}. We will analyse various interesting cases 
in detail below.

First, an interesting subfamily of restricted states is obtained if we
require that the ghost-spins are correlated, \ie\ with the ghost-spins
being identical in each basis state: then the only allowed states are
\be\label{12gsCorrelated}
|\psi\rangle = \psi^{+,++} |+\rangle |++\rangle 
+ \psi^{+,--} |+\rangle |--\rangle 
+\ \psi^{-,++} |-\rangle |++\rangle + \psi^{-,--} |-\rangle |--\rangle\ .
\ee
Since there is an even number of minus signs, this entire subfamily of 
states is manifestly positive norm, from (\ref{psi12mostgenNorm}). In 
other words, we have excluded all negative norm states and thus 
we recover positivity of entanglement entropy manifestly (as can be 
verified from (\ref{rhoA12mostgen})).

More generally, there are minus signs in $\rho_A$. To explore the 
possibilities for positive norm states giving $S_A>0$, consider the 
relatively simple but instructive subfamily of states
\be\label{psi12gen}
|\psi\rangle = \psi^{+,++} |+\rangle |++\rangle 
+ \psi^{+,+-} |+\rangle |+-\rangle + \psi^{-,-+} |-\rangle |-+\rangle 
+ \psi^{-,--} |-\rangle |--\rangle
\ee
(which in general are not product states). These have norm
\be
\langle\psi|\psi\rangle = |\psi^{+,++}|^2 - |\psi^{+,+-}|^2
- |\psi^{-,-+}|^2 + |\psi^{-,--}|^2
\ee
which are negative if
\be
|\psi^{+,++}|^2 + |\psi^{-,--}|^2 < |\psi^{+,+-}|^2 + |\psi^{-,-+}|^2\ ,
\ee
somewhat similar to the two ghost-spins system reviewed earlier. The 
states (\ref{psi12gen}) have sufficiently many 
parameters while restricting to a subfamily with $\rho_A$ 
diagonal (thus simplifying $\log\rho_A$ as well) and 
$(\rho_A)^{++}, (\rho_A)^{--} > 0$ for positive norm states.
Explicitly, the reduced density matrix (\ref{rhoA12-1}) 
becomes
\bea
(\rho_A)^{++} = |\psi^{+,++}|^2 - |\psi^{+,+-}|^2\ , & \quad &
(\rho_A)^{+-} = 0\ ,\nonumber\\
(\rho_A)^{-+} = 0\ , & \quad &
(\rho_A)^{--} = -|\psi^{-,-+}|^2 + |\psi^{-,--}|^2\ .
\eea
Note that $tr\rho_A = g_{ik} (\rho_A)^{ik} = (\rho_A)^{++} + (\rho_A)^{--}$ 
and satisfies $tr\rho_A = tr\rho = \langle\psi|\psi\rangle$. Since 
the remaining spin has positive definite metric $g_{ij}$, we have 
the entanglement entropy
\be
S_A = -g_{ij} (\rho_A\log\rho_A)^{ij} = - (\rho_A\log\rho_A)^{++} 
- (\rho_A\log\rho_A)^{--}\ .
\ee
For states with positive/negative norm, we can normalize as 
\be\label{psi12-norm}
\langle\psi|\psi\rangle = |\psi^{+,++}|^2 - |\psi^{+,+-}|^2
- |\psi^{-,-+}|^2 + |\psi^{-,--}|^2 = \pm 1\ ,
\ee
so that the entanglement entropy becomes
\be\label{rhoA12-x}
\begin{split}
  |\psi^{+,++}|^2 - |\psi^{+,+-}|^2 \equiv x ,\qquad 
\langle\psi|\psi\rangle = x + (\pm 1-x)\ ; \\
(\rho_A)^{+,+} =  x ,\qquad (\rho_A)^{-,-} = \pm 1 - x ,\quad \\
\qquad\qquad S_A = - x\log x - (\pm 1 - x) \log (\pm 1 - x)\ .
\end{split}
\ee
{\bf Positive norm:}\ \ Now if $x>0$, then $(1-x)>0$ also from 
(\ref{psi12-norm}), (\ref{rhoA12-x}), implying $0<x<1$ and $(\rho_A)$ 
is positive definite.  This gives
\be
S_A = -x\log x - (1-x) \log(1-x) > 0\ .
\ee
If $x<0$, then $(1-x)>0$, giving $(\rho_A)^{++}<0,\ (\rho_A)^{--}>0$, 
with 
\be
S_A = |x|\log |x| - (1+|x|)\log(1+|x|) + i\pi |x|\ ,
\ee
where the real part is negative which shows anti-correlation with the
norm.   In addition there is an imaginary part which depends linearly
on $x$. 

This behaviour of $S_A$ can be interpreted as follows. Choosing $x>0$
means we assign higher probability to getting the $|++\rangle$
ghost-spin state than the $|+-\rangle$ state (and likewise the
$|--\rangle$ versus $|-+\rangle$). Since the $|++\rangle$ and $|--\rangle$
are positive norm states (in the sense of (\ref{12gsCorrelated}) with
correlated ghost-spins), $x>0$ corresponds to the component of the 
Hilbert space continuously connected to the correlated ghost-spin sector
(which contains only $|++\rangle$ and $|--\rangle$ ghost-spin basis
states).  Equivalently starting with the correlated ghost-spin sector
of the Hilbert space, small deformations of the state vector by
turning on small $|+-\rangle$ components (or $|-+\rangle$) are still
positive norm only if $x>0$.  For $x<0$, this feature does not exist
simply because the corresponding state is not continuously connected
to the positive norm correlated ghost-spin sector.
 
\noindent {\bf Negative norm:}\ \ Now $\rho_A$ is negative definite if\
$x<0$ and $(-1-x)<0$, so that\ $0<|x|<1$, giving
\be
S_A = |x| \log |x| + (1-|x|) \log (1-|x|) + i\pi\ ,
\ee
with a negative real part since the $\log$s are negative, and a constant 
imaginary part.\\
With $x<-1$, we have $(\rho_A)^{++}<0$ and $(\rho_A)^{--}=-1+|x|>0$, 
which gives 
\be
S_A = |x| \log |x|  - (|x|-1) \log (|x|-1) + i\pi |x|\ ,
\ee
where the real part is positive but the imaginary part is not constant
anymore.  Likewise when we have $(\rho_A)^{++}>0,\ (\rho_A)^{--}<0$,
which corresponds to $x>0$ and $-1-x<0$, we get
\be
S_A = -x\log x + (1+x) \log (1+x) + i\pi (1+x)\ ,
\ee
where the real part is again positive.  Finally the choice
$(\rho_A)^{++}>0,\ (\rho_A)^{--}>0$, gives $x>0,\ -1-x>0$,  which is
clearly not possible.

To investigate other possibilities, let us consider restricting to a
diagonal $\rho_A$,  this corresponds to setting $(\rho_A)^{+-},\
(\rho_A)^{-+}=0$.  This gives rise to the reduced density matrix
corresponding to yet another subfamily of entangled states. Setting
to zero the off-diagonal terms in the reduced density matrix 
(\ref{rhoA12mostgen}) gives
\be
(\psi^*)^{-,+-} = {1\over \psi^{+,+-}} (\psi^{+,++}(\psi^*)^{-,++} 
- \psi^{+,-+}(\psi^*)^{-,-+} + \psi^{+,--}(\psi^*)^{-,--})
\ee
and a complex conjugated condition.  Using these conditions the norm
of the state as well as the diagonal components of the reduced density
matrix can be written as
\bea
\langle\psi|\psi\rangle =&& |\psi^{+,++}|^2 - |\psi^{+,+-}|^2
- |\psi^{+,-+}|^2 + |\psi^{+,--}|^2 + |\psi^{-,++}|^2 - |\psi^{-,-+}|^2 
+ |\psi^{-,--}|^2 \nonumber\\
&& - {1\over|\psi^{+,+-}|^2} |(\psi^{+,++}(\psi^*)^{-,++} 
- \psi^{+,-+}(\psi^*)^{-,-+} + \psi^{+,--}(\psi^*)^{-,--})|^2
\eea
and
\bea
&& (\rho_A)^{++} = |\psi^{+,++}|^2 - |\psi^{+,+-}|^2 - |\psi^{+,-+}|^2
+ |\psi^{+,--}|^2\ ,\nonumber\\
&& \qquad\qquad  (\rho_A)^{--} = |\psi^{-,++}|^2 - |\psi^{-,+-}|^2 
- |\psi^{-,-+}|^2 + |\psi^{-,--}|^2\ .\
\eea
This form of $\rho_A$ is again quite flexible, \ie\ there are
subregions in the Hilbert space where positive norm states give
$S_A>0$, which is along the lines of (\ref{rhoA12-x}).  The resulting
analysis therefore follows the same pattern.  Although the generic
state does not give positive $S_A$ for positive norm states unless we
impose further restrictions, if we allow all such basis states, then
we can always find subfamilies of states where $S_A>0$ for positive
norm states.

\section{Spins entangled with one ghost-spin}\label{sec:spins-entangled}

In this section we will consider entangled systems containing one
ghost-spin.  We will start with one spin entangled with one ghost-spin
and later generalize it to multiple spins entangled with one ghost-spin.

\subsection{One spin entangled with one ghost-spin}\label{sec:one-spin-entangled}

Here we will demonstrate that whenever we have one ghost-spin
entangled with an ordinary spin, it is not possible to find an
entangled state which has positive entanglement entropy for positive
norm states and negative entanglement entropy for negative norm states
after tracing over the ghost-spin.  Although we will use one spin and
one ghost-spin system, it is easy to see that the conclusion is
independent of the number of spins in the system (after tracing out
the ghost-spin) as we will see in the next section. The reason is that
the outcome completely depends on the ghost-spin system.

A point to note here is that the entanglement entropy, no matter
whether we have positive norm states or negative norm states, is
necessarily a complex quantity with non-constant imaginary part.  This
is to be contrasted with the disentangled system where we have complex
entanglement entropy for negative norm states but the imaginary part
was constant.  We also find that the real part of the entanglement
entropy is anti-correlated with the norm, \ie\ positive norm states
have negative definite real part of the entropy and vice versa.

A generic state for one spin and one ghost-spin system is
\be\label{psi11}
|\psi\rangle = \psi^{i,\al} |i\rangle |\al\rangle  
\ee
where $i=\pm$ refers to the spin index while $\al=\pm$ refers to the 
ghost-spin index. Then the norm is
\be\label{psi11norm}
\begin{split}
  \langle\psi|\psi\rangle 
&= g_{ij} \gamma_{\al\beta} \psi^{i,\al}(\psi^*)^{j,\beta}
= \sum_{i,\al} \gamma_{\al\al}\psi^{i,\al}(\psi^*)^{i,\al} \\
&= |\psi^{+,+}|^2 - |\psi^{+,-}|^2 + |\psi^{-,+}|^2 - |\psi^{-,-}|^2 = \pm 1\ ,
\end{split}
\ee
where the normalization $\pm 1$ refers to positive/negative norm states 
respectively. This is negative norm if\ 
$|\psi^{+,+}|^2 - |\psi^{+,-}|^2 + |\psi^{-,+}|^2 - |\psi^{-,-}|^2<0$.

From sec.~\ref{sec:rdm-gs}\ (see (\ref{rdm-gs}), (\ref{rdm-gs-rho}), 
(\ref{rdm-gs-rhoGen})), the reduced density matrix obtained by tracing 
out the ghost-spin in the present case is
\be\label{rhoA11-1}
(\rho_A)^{ik} = \gamma_{\al\beta} \psi^{i,\al} (\psi^*)^{k,\beta} 
= \gamma_{\al\al} \psi^{i,\al} (\psi^*)^{k,\al}
\ee
\bea\label{rhoA11-2}
\Rightarrow\qquad\qquad (\rho_A)^{++} = |\psi^{+,+}|^2 - |\psi^{+,-}|^2\ , 
& \quad & 
(\rho_A)^{+-} = \psi^{+,+} (\psi^*)^{-,+} - \psi^{+,-} (\psi^*)^{-,-}\ ,
\nonumber\\
(\rho_A)^{-+} = \psi^{-,+} (\psi^*)^{+,+} - \psi^{-,-} (\psi^*)^{+,-}\ ,& 
\quad &
(\rho_A)^{--} = |\psi^{-,+}|^2 - |\psi^{-,-}|^2\ .
\eea
This is identical to the case of the two ghost-spin system, except that 
$\rho_A$ is now contracted with the positive definite spin metric.\
The mixed-index reduced density matrix is obtained by raising an index 
with the (positive definite) spin metric, giving\
$(\rho_A)_i^k = g_{ij} (\rho_A)^{jk}\ \Rightarrow\ 
(\rho_A)^+_+ = (\rho_A)_{++} , \ (\rho_A)^-_- = (\rho_A)_{--}$.
Focussing on the subfamily with\ $(\rho_A)_{+-}=0$ such that the
reduced density matrix is diagonal, we have\ (see 
Appendix~\ref{sec:diag-dens-matr} for non-diagonal $\rho_A$)\ 
\bea
&& (\psi^*)^{-,+} = {\psi^{+,-} (\psi^*)^{-,-}\over \psi^{+,+} }\ ,\qquad\qquad
{1\over |\psi^{+,+}|^2} (|\psi^{+,+}|^2 - |\psi^{+,-}|^2) 
(|\psi^{+,+}|^2 - |\psi^{-,-}|^2) = \pm 1\ ,\nonumber\\
&& (\rho_A)^{++} = |\psi^{+,+}|^2 - |\psi^{+,-}|^2\ ,\qquad
(\rho_A)^{--} = -{|\psi^{-,-}|^2\over |\psi^{+,+}|^2} 
(|\psi^{+,+}|^2 - |\psi^{+,-}|^2)\ .
\eea
We see that $(\rho_A)^{--}$ necessarily has sign opposite to $(\rho_A)^{++}$, 
so all states, including positive norm states, necessarily have a 
negative eigenvalue. Simplifying gives
\be\label{eq:rdm1+1}
(\rho_A)^{++} = \pm {|\psi^{+,+}|^2\over |\psi^{+,+}|^2 - |\psi^{-,-}|^2} 
\equiv \pm x\ ,
\qquad
(\rho_A)^{--} = \mp {|\psi^{-,-}|^2\over |\psi^{+,+}|^2 - |\psi^{-,-}|^2} 
= \pm (1-x)\ .
\ee
Due to the relative sign between the two terms in the denominator we end
up with $|x|>1$. Then tracing over the remaining spin using its positive 
definite metric gives the entanglement entropy
\bea
S_A &=& -g_{ij} (\rho_A \log\rho_A)^{ij} 
=  -(\rho_A\log\rho_A)^+_+ - (\rho_A\log\rho_A)^-_- \nonumber\\
&=& - (\pm x) \log (\pm x) - (\pm (1-x)) \log (\pm (1-x))\ .
\eea
Since $1-x$ is necessarily negative for positive norm states, we have an
imaginary component in the entanglement entropy,
\be
S_A = -x\log x + (x-1)\log (x-1) + i\pi (x-1)\ .
\ee
Note that the real part of the entropy is negative and the imaginary
part is $x$-dependent.  In other words, $S_A$ is not positive for
positive norm states.  It turns out that this is a generic feature of
this system.  For example we could try to consider restricted cases, 
{\em i.e.} we can fix some of the parameters to see if special entangled
states can exhibit positivity of entanglement entropy for positive
norm states.  As will be illustrated in the example below, one
generically fails to ensure positivity and reality of the entanglement
entropy. 

\noindent 
\emph{Example:}\ \ taking $\psi^{+,-}=0$ as a 
special case gives\ $|\psi\rangle = \psi^{+,+} |+\rangle|+\rangle + 
\psi^{-,-} |-\rangle|-\rangle$,\ and
\begin{equation}
  \label{eq:14}
  \begin{split}
    & |\psi^{+,+}|^2 - |\psi^{-,-}|^2 = \pm 1\ ,\ \ 
    (\log\rho_A)^+_+ = \log(|\psi^{+,+}|^2)\ ,\quad 
    (\log\rho_A)^-_- = \log(-|\psi^{-,-}|^2)\ ,\\
    & 
    S_A = -|\psi^{+,+}|^2 \log \big(|\psi^{+,+}|^2\big) + 
    |\psi^{-,-}|^2 \log \big(|\psi^{-,-}|^2\big) + |\psi^{-,-}|^2 (i\pi)\ .
  \end{split}
\end{equation}
We see that this now depends on whether the state is positive or negative 
norm, which are normalized respectively as\ $\pm 1$.\ For positive norm 
states, \ we take $|\psi^{-,-}|^2 = |\psi^{+,+}|^2 - 1$,\ so
\begin{equation}
  \label{eq:15}
  S_A = -x\log x + (x-1)\log(x-1) + (x-1)(i\pi)\ ,\qquad x=|\psi^{+,+}|^2 ,
\qquad  1\leq x<\infty\ .
\end{equation}
The real part of $S_A$ is always negative definite, although this 
is a positive norm state: also there is an imaginary part. 
For negative norm states, \ we take $|\psi^{-,-}|^2 = |\psi^{+,+}|^2 + 1$,\ 
giving
\begin{equation}
  \label{eq:16}
  S_A = -x\log x + (x+1)\log(x+1) + (x+1)(i\pi)\ ,\qquad x=|\psi^{+,+}|^2 ,
\qquad  0\leq x<\infty\ .
\end{equation}
In this case, the real part of $S_A$ is positive definite, although 
this is a negative norm state.

\vspace{2mm}

Tracing over the spin first gives 
$(\rho_A)^{\alpha\beta} = g_{ij} \psi^{i,\alpha} (\psi^*)^{j,\beta}$
which has no negative  
signs since $g_{ij}$ is positive definite. But the mixed-index reduced 
density matrix is $(\rho_A)_\alpha^\beta = \gamma_{\alpha\delta}(\rho_A)^{\delta\beta}$,
identical to the one so far. In particular for the example above, 
with $\psi^{+-}=0$, we have\ 
$(\rho_A)^{++} = |\psi^{+,+}|^2,\ (\rho_A)^{--} = |\psi^{-,-}|^2$, 
while\ $(\rho_A)^+_+ = |\psi^{+,+}|^2,\ (\rho_A)^-_- = -|\psi^{-,-}|^2$.\ 
The entanglement entropy is obtained by contracting with the metric 
$\gamma_{\alpha\beta}$ of the remaining ghost-spin, giving\ 
$S_A = -\gamma_{\alpha\beta} (\rho_A\log\rho_A)^{\alpha\beta} 
= -(\rho_A)^+_+ (\log\rho_A)^+_+ - (\rho_A)^-_- (\log\rho_A)^-_-$ as 
before.

\vspace{2mm}

Given the result so far, it is worth asking if any other contraction
scheme for the reduced density matrix yields something useful, although 
this is not expected to satisfy the physical conditions: let us 
therefore consider the analog of (\ref{rdm12-2}) in 
sec.~\ref{sec:rdm-gs}. In the present case, tracing over the ghost-spin 
would correspond to
\be
(\rho_A)^{ik} = \gamma_{\al\beta} \psi^{i,\al} (\psi^*)^{\beta,k} 
= \gamma_{\al\al} \psi^{i,\al} (\psi^*)^{\al,k}\ .
\ee
See also the discussion around eq.\eqref{rdm12-2} for a similar scheme for
the case of one spin and two ghost-spins. The components of the
reduced density matrix are
\bea\label{rhoA11-3}
(\rho_A)^{++} = |\psi^{+,+}|^2 - \psi^{+,-} (\psi^*)^{-,+}\ , & \qquad & 
(\rho_A)^{+-} = \psi^{+,+} (\psi^*)^{+,-} - \psi^{+,-} (\psi^*)^{-,-}\ ,
\nonumber\\
(\rho_A)^{-+} = \psi^{-,+} (\psi^*)^{+,+} - \psi^{-,-} (\psi^*)^{-,+}\ ,
& \qquad &
(\rho_A)^{--} = \psi^{-,+}(\psi^*)^{+,-} - |\psi^{-,-}|^2\ .
\eea
Requiring the hermiticity condition on $\rho^A$ then implies 
$\psi^{+,-}=\psi^{-,+}$. Substituting this condition back into the form 
of the reduced density matrix gives
\bea\label{rhoA11-4}
(\rho_A)^{++} = |\psi^{+,+}|^2 - |\psi^{+,-}|^2\ , & \qquad & 
(\rho_A)^{+-} = \psi^{+,+} (\psi^*)^{+,-} - \psi^{+,-} (\psi^*)^{-,-}\ ,
\nonumber\\
(\rho_A)^{-+} = \psi^{+,-} (\psi^*)^{+,+} - \psi^{-,-} (\psi^*)^{+,-}\ ,
& \qquad &
(\rho_A)^{--} = |\psi^{+,-}|^2 - |\psi^{-,-}|^2\ .
\eea
This is diagonal if $|\psi^{+,+}|^2=|\psi^{-,-}|^2$, however $\rho_A$
is again not positive definite.  As a result, this different
contraction rule also does not help improve the situation (although
this was simply a technical exercise, not accounting for the physical
conditions).  It is just as well that the conundrum of the
anti-correlation of the sign of the entanglement entropy with the norm
is unaffected by this alternate contraction scheme.  This is because a
single ghost-spin system is analogous to a gauge theory with a single
set of indefinite norm states.  We know that we need a second set of
indefinite norm fields, namely the ghost fields, to effectively impose
the restriction to the physical subspace.

\subsection{Multiple spins entangled with one ghost-spin}\label{sec:k+1}

We will now show that the results obtained for the system studied in
the previous subsection \ref{sec:one-spin-entangled} can be easily
extended to an arbitrary number of spins entangled with one ghost-spin.
This establishes that the anti-correlation between the norm of the
state and sign of the real part of the entanglement entropy is an
effect entirely due to tracing over the single ghost-spin degree of
freedom. 

Let us consider a system with $k$ spins entangled with a single
ghost-spin.  As usual we will denote the ghost spin by $\pm$;\ however,
to denote the $k$-tuple of ordinary spins, we will use indices $I, J,
\cdots$.  They run over $2^k$ possible configurations of $k$ spins.  A
state with spin configuration $I$ and ghost spin $+$ is denoted as
$|I,+\rangle$.  Using this notation, we can write the reduced density
matrix after tracing over the ghost-spin degrees of freedom as
\begin{equation}
  \label{eq:7}
  \begin{split}
    (\rho_A)^{I,J} &= \psi^{I,+}(\psi^{J,+})^* - (\psi^{I,-})^*\psi^{J,-}  \\
    (\rho_A)^{I,I} &= |\psi^{I,+}|^2 - |\psi^{I,-}|^2\ .
  \end{split}
\end{equation}
To illustrate the point, let us restrict to the diagonal form of the
reduced density matrix.  We will see in a moment that there is no loss
of generality in doing this.  Setting the off diagonal components of the
density matrix to zero gives the condition
\begin{equation}
  \label{eq:8}
  \psi^{I,+} = \frac{(\psi^{I,-})^*\psi^{J,-}}{(\psi^{J,+})^*}\ .
\end{equation}
We can now relate the diagonal components of $\rho_A$ using the condition
\eqref{eq:8}, obtaining
\begin{equation}
  \label{eq:9}
  (\rho_A)^{J,J} = -(\rho_A)^{I,I}  \frac{|\psi^{J,+}|^2}{|\psi^{I,-}|^2}\ .
\end{equation}
This relation is the multi-spin generalization of \eqref{eq:rdm1+1} 
and is valid for any pair $(I,J)$. Therefore, further analysis of this 
system has similarities with that in section \ref{sec:one-spin-entangled}. 
In particular we see that as long as we have entangled states with a 
single ghost-spin and we trace over it then, in general, we always end 
up obtaining negative eigenvalues in the reduced density matrix $\rho_A$.
Thus positive norm states do not lead to positive entanglement entropy. 
This result is independent of the number of ordinary spins in the
entangled state.

We will now comment on the choice of the reduced density matrix.
Suppose we are considering a conventional spin system.  Since the
metric on this space is positive definite, the entropy of entangled
states is always positive definite, no matter what choice of basis we
select to denote these spin states.  Let us now consider this spin
system entangled with a ghost-spin.  Since our aim is to trace over
the ghost-spin and write down the reduced density matrix for the
conventional spins, the choice of basis in the conventional spin space
does not affect the form of the reduced density matrix.

\section{Multi-ghost-spin systems}\label{sec:multi-ghost-spin}

In this section we will consider multiple entangled ghost-spin
systems.  As a warm up, we will first look at the three entangled
ghost-spins system and then generalize it to multiple entangled
ghost-spins.  We find that in a certain class of entangled states it
is easy to distinguish entangled states of an even number of
ghost-spins from those involving an odd number of ghost-spins.  When
the number of ghost-spins is even, then after tracing over all the
ghost-spins except one, we get manifestly positive definite
entanglement entropy.  However, when the number of ghost-spins is odd,
then following the same procedure of tracing over all ghost-spins
except one gives a negative definite real part of the entanglement
entropy for positive norm states.  We will also consider entangling
this multi-ghost-spin system with one spin.  After tracing over all
ghost-spins, the resulting entanglement entropy exhibits the same odd
vs. even distinction as the pure multi-ghost-spin system.

\subsection{Three ghost-spins}

We will begin with a system of three entangled ghost-spins.
A generic state and its norm are
\be
|\psi\rangle = \psi^{\al\beta\gamma} |\al\beta\gamma\rangle\ ,\qquad
\langle\psi|\psi\rangle = \gamma_{\al\delta} \gamma_{\beta\sigma} 
\gamma_{\gamma\rho}\psi^{\al\beta\gamma}(\psi^*)^{\delta\sigma\rho}\ .
\ee
Explicitly writing the most general state, we have
\be\label{psi03mostgen}
\begin{split}
  |\psi\rangle =& \ \psi^{+++} |+++\rangle 
+ \psi^{++-} |++-\rangle + \psi^{+-+} |+-+\rangle
+ \psi^{+--} |+--\rangle  \ \\
&\ \ +\ \psi^{-++} |-++\rangle + \psi^{-+-} |-+-\rangle 
+ \psi^{--+} |--+\rangle + \psi^{---} |---\rangle
\end{split}
\ee
with norm
\be\label{psi03mostgenNorm}
\begin{split}
  \langle\psi|\psi\rangle =& \ |\psi^{+++}|^2 - |\psi^{++-}|^2
- |\psi^{+-+}|^2 + |\psi^{+--}|^2 \\ &\ - |\psi^{-++}|^2 + |\psi^{-+-}|^2
+ |\psi^{--+}|^2 - |\psi^{---}|^2\ .
\end{split}
\ee
The reduced density matrix 
$(\rho_A)^{\al\delta} = \gamma_{\beta\sigma} \gamma_{\gamma\rho} 
\psi^{\al\beta\gamma} (\psi^*)^{\delta\sigma\rho} = 
\gamma_{\beta\beta} \gamma_{\rho\rho} \psi^{\al\beta\rho} (\psi^*)^{\delta\beta\rho}$ 
for the last ghost-spin after tracing over two ghost-spins is
\bea\label{rhoA03mostgen}
 (\rho_A)^{++} &=& |\psi^{+++}|^2 - |\psi^{++-}|^2 - |\psi^{+-+}|^2
+ |\psi^{+--}|^2\ , \qquad \nonumber\\
(\rho_A)^{+-} &=& \psi^{+++}(\psi^*)^{-++} - \psi^{++-}(\psi^*)^{-+-}
- \psi^{+-+}(\psi^*)^{--+} + \psi^{+--}(\psi^*)^{---}\ ,\nonumber\\
(\rho_A)^{-+} &=& \psi^{-++}(\psi^*)^{+++} - \psi^{-+-}(\psi^*)^{++-}
- \psi^{--+}(\psi^*)^{+-+} + \psi^{---}(\psi^*)^{+--}\ , \qquad \\
(\rho_A)^{--} &=& |\psi^{-++}|^2 - |\psi^{-+-}|^2 
- |\psi^{--+}|^2 + |\psi^{---}|^2\ .\nonumber
\eea
Consider the relatively simple but instructive subfamily of states
\be\label{psi03gen}
|\psi\rangle = \psi^{+++} |+++\rangle + \psi^{++-} |++-\rangle 
+ \psi^{--+} |--+\rangle + \psi^{---} |---\rangle
\ee
(which in general are not product states) with normalized norm
\be\label{psi03gen-norm}
\langle\psi|\psi\rangle = |\psi^{+++}|^2 - |\psi^{++-}|^2
+ |\psi^{--+}|^2 - |\psi^{---}|^2 = \pm 1\ .
\ee
The reduced density matrix becomes
\bea\label{psi03gen-rhoA}
(\rho_A)^{++} = |\psi^{+++}|^2 - |\psi^{++-}|^2\ , & \quad &
(\rho_A)^{+-} = 0\ ,\nonumber\\
(\rho_A)^{-+} = 0\ , & \quad &
(\rho_A)^{--} = -|\psi^{--+}|^2 + |\psi^{---}|^2\ ,
\eea
and the mixed-index $(\rho_A)_\alpha^\kappa = \gamma_{\alpha\beta} (\rho_A)^{\beta\kappa}$ is\
$(\rho_A)^+_+ = (\rho_A)^{++},\ (\rho_A)^-_- = -(\rho_A)^{--}$, so that
$tr\rho_A = (\rho_A)^+_+ + (\rho_A)^-_- = \langle\psi|\psi\rangle$.
Also $(\log\rho_A)^+_+ = \log((\rho_A)^+_+)$ etc.
So the entanglement entropy becomes\
$S_A = - (\rho_A)^+_+ (\log\rho_A)^+_+ - (\rho_A)^-_- (\log\rho_A)^-_-$. 
We then have (similar to (\ref{rhoA12-x}) in the case of one spin and 
two ghost-spins)
\be\label{rhoA03-x}
\begin{split}
  |\psi^{+++}|^2 - |\psi^{++-}|^2 \equiv x ,\qquad 
\langle\psi|\psi\rangle = x + (\pm 1-x)\ ;\\
(\rho_A)^+_+ =  x ,\qquad
(\rho_A)^-_- = \pm 1 - x , \quad \\
\qquad\qquad S_A = - x\log x - (\pm 1 - x) \log (\pm 1 - x)\ .
\end{split}
\ee
As in the discussion following (\ref{rhoA12-x}), we can see that 
positive norm states with $x>0$ have $S_A>0$ but when $x<0$ we obtain 
$Re(S_A)<0$ and $Im(S_A)\neq 0$. However unlike in that case, 
there is no correlated ghost-spin subsector here so no reason to 
restrict to $x>0$ states.

\subsection{Multiple ghost-spins}

Consider a system of $n$ ghost-spins and the entangled state (and its 
norm)
\be\label{++++----}
|\psi\rangle = \psi^{++\ldots} |++\ldots\rangle 
+ \psi^{--\ldots} |--\ldots\rangle\ ,\qquad
\langle\psi|\psi\rangle = |\psi^{++\ldots}|^2 + (-1)^n |\psi^{--\ldots}|^2\ .
\ee
This state is a linear combination of a state with all $+$ and another 
with all $-$.
The mixed-index reduced density matrix for a subsystem comprising a 
single ghost-spin after tracing over the remaining $n-1$ ghost-spins, 
and associated entanglement entropy are
\bea
&& (\rho_A)^+_+ = (\rho_A)^{++} = |\psi^{++\ldots}|^2\ ,\qquad
(\rho_A)^-_- = -(\rho_A)^{--} = (-1)^n |\psi^{--\ldots}|^2\ ,\nonumber\\
&& S_A = - (\rho_A)^+_+ (\log\rho_A)^+_+ - (\rho_A)^-_- (\log\rho_A)^-_-\ .
\eea
For $n$ even, the state is clearly positive norm and so has manifestly 
positive $\rho_A$ and thus positive entanglement entropy. For $n$ odd 
however, we have\ $|\psi^{++\ldots}|^2 - |\psi^{--\ldots}|^2 = \pm 1$ for 
normalized positive/negative norm states, giving
\be
S_A = -|\psi^{++\ldots}|^2 \log \big(|\psi^{++\ldots}|^2\big) + 
|\psi^{--\ldots}|^2 \log \big(|\psi^{--\ldots}|^2\big) 
+ |\psi^{--\ldots}|^2 (i\pi)\ ,
\ee
very similar to the case of one spin entangled with one ghost-spin 
(see \eg\ eqs.\eqref{eq:14}, \eqref{eq:15}, and \eqref{eq:16}). In 
particular, for positive norm states, $S_A$ has a negative definite
real part (and an imaginary part) while for negative norm states
$Re(S_A)>0$. The states (\ref{++++----}) always exist and have this
counter-intuitive structure for odd numbers of ghost-spins.

In the above, we could implicitly regard the $n$ ghost-spins as a
``ghost-spin-chain'' in a 1-dimensional space with the ghost-spins
located at lattice sites. Then the single ghost-spin comprises a
subchain whose entanglement with the rest of the chain has the above
structure.


\subsection{One spin entangled with multiple ghost-spins}
\label{sec:one-spin-entangled-3}

We will generalize the case studied in the last subsection by coupling
it to one ordinary spin.  As we will see, the outcome is identical in
the sense that there is a clear distinction between states with an odd
number of ghost-spins and those with an even number of ghost-spins.
The reduced density matrix after tracing over all ghost-spins is
positive definite for even number of ghost-spins which contains
positive norm states.  In the case of odd numbers of ghost-spins, we
exhibit simple entangled states which always exhibit anticorrelation
between their norm and the sign of the real part of the entanglement
entropy.

Consider a system consisting of a spin and $n$ ghost-spins.  We will
look at a simple entangled state which is a straightforward
generalization of the $n$ ghost-spin state in (\ref{++++----}).  We 
denote the state with the first index representing the spin and the 
rest representing the ghost-spins,
\begin{equation}
  \label{eq:17}
  \begin{split}
  |\psi\rangle_{(1,n)} &= \psi^{+,++\ldots} |+,++\ldots\rangle 
+ \psi^{-,--\ldots} |-,--\ldots\rangle\ ,\\
_{(1,n)}\langle\psi|\psi\rangle_{(1,n)} &= |\psi^{+,++\ldots}|^2 + (-1)^n
|\psi^{-,--\ldots}|^2\ . 
  \end{split}
\end{equation}
The reduced density matrix in the ordinary spin sector with mixed
indices obtained after tracing over all $n$ ghost-spins has the form
\begin{equation}
  \label{eq:18}
   (\rho_A)^+_+ = (\rho_A)^{++} = |\psi^{+,++\ldots}|^2\ ,\qquad
(\rho_A)^-_- = (\rho_A)^{--} = (-1)^n |\psi^{-,--\ldots}|^2\ .
\end{equation}
The entanglement entropy for the single spin subsystem is
\begin{equation}
  \label{eq:19}
  S_A = - (\rho_A)^+_+ (\log\rho_A)^+_+ - (\rho_A)^-_-
  (\log\rho_A)^-_-\ .
\end{equation}
It is obvious from the expressions for the components of the reduced
density matrix \eqref{eq:18} that for $n$ even the state has positive
norm and the entanglement entropy is positive definite.  However, when
$n$ is odd then we are back to the situation where positive norm does
not lead to positive entanglement entropy.  This is identical to the
situation encountered in the system with a single spin entangled with 
a single ghost-spin. This conclusion also applies if there are multiple 
spins instead of one.

We therefore conclude that the multi-ghost-spin systems fall into two
categories.  Whereas the even number of ghost-spins case gives rise to
positive norm states and positive entanglement entropy, the odd number
of ghost-spins case always contain states such as (\ref{++++----}),
(\ref{eq:17}), which exhibit the unphysical anticorrelation between
the norm and the sign of the real part of the entanglement 
entropy\footnote{although there are also states with positive norm 
and positive entanglement, \eg\ (\ref{psi03gen}), 
(\ref{psi03gen-norm}), (\ref{psi03gen-rhoA}), (\ref{rhoA03-x}).}
that we first encountered in the case of the single ghost-spin system.
This fits well with our interpretation that an even number of indefinite
norm states is needed to get a sensible reduction to the definite norm
subsector of the theory.  In this sense, the odd number of ghost-spins
systems are analogous to partial gauge fixed or gauge unfixed systems.

\section{Discussion}

We have studied patterns of quantum entanglement in spin \& 
ghost-spin systems. When the spins and ghost-spins are disentangled (both
sectors possibly entangled within themselves), the reduced density
matrix obtained by tracing out the ghost-spins leads to positive
entanglement entropy for positive norm states.  Negative norm states
give rise to entanglement entropy with a negative real part and a
constant imaginary part. However, for entangled spins and ghost-spins,
the entanglement patterns are richer. For even numbers of
ghost-spins, there are always subsectors of the Hilbert space where
positive norm states give positive entanglement entropy. For odd
numbers of ghost-spins, we have seen the existence of positive norm
states which always have negative real part for entanglement
entropy. These toy models in a sense contain only entanglement
information: we have not utitized any description of time evolution
and dynamics. It would be interesting to further explore dynamical
models which lead to the toy models here.  It would also be
interesting to explore inter-relations of these models with recent
studies of entanglement in gauge theories \eg\
\cite{Casini:2013rba,Ghosh:2015iwa,Soni:2015yga,Casini:2015dsg,Soni:2016ogt}.
In this regard, it is interesting to note \cite{Kabat:1995eq} who 
point out the necessity of ghost fields to account for entanglement 
in gauge theories: see also related discussions more recently in \eg\ 
\cite{Donnelly:2015hxa,Harlow:2015lma}.

A related obvious question in the present context has to do with how
precisely the physical positive norm subspace arises in the full
theory containing the negative norm sectors. It would appear that such
a truncation must dovetail with a better understanding of the partial
trace in the reduced density matrix over the extended Hilbert space
and the gauge fixed theory including the ghost sector arising from
gauge fixing.  In general understanding how the physical subspace
arises is unclear within the present work, possibly admitting a clear
answer in the context of toy models with dynamics \eg\
\cite{Jatkar:2017jwz} and ongoing investigations on the role of a BRST
symmetry and associated cohomology. It would appear that the latter
will provide a truncation to a physical subspace which is entirely
positive norm, thereby leading to positive definite reduced density
matrices and positive entanglement (which can be expected to satisfy
known universal properties such as strong subadditivity). In general
this may not be as simple as truncating to \eg\ correlated ghost-spin
subsectors (although that subsector is indeed entirely positive norm
in simple toy models). We hope to clarify some of these issues in
future work.

The motivation for defining ``ghost-spins'' in \cite{Narayan:2016xwq}
arose from $dS/CFT$, which, although not directly relevant to the
present context, is useful to review briefly.  Certain generalizations
of gauge/gravity duality to de Sitter space or $dS/CFT$
\cite{Strominger:2001pn,Witten:2001kn,Maldacena:2002vr} conjecture
that de Sitter space is dual to a hypothetical Euclidean non-unitary
CFT that lives on the future boundary ${\cal I}^+$. The late-time
wavefunction of the universe $\Psi_{dS}$ with appropriate boundary
conditions is equated with the dual CFT partition function $Z_{CFT}$
\cite{Maldacena:2002vr}, which is a useful way to organize de Sitter
perturbations (independent of the actual existence of the CFT). The
dual CFT$_d$ energy-momentum tensor correlator $\langle T T\rangle$ in
a semiclassical approximation $\Psi\sim e^{iS}$ reveals central charge
coefficients ${\cal C}_d\sim i^{1-d}{R_{dS}^{d-1}\over G_{d+1}}$ in
$dS_{d+1}$, real and negative in $dS_4$, and pure imaginary in $dS_3,
dS_5$ etc (effectively analytic continuations from $AdS/CFT$). 
$dS_4/CFT_3$ is thus reminiscent of ghost-like non-unitary theories. 
In \cite{Anninos:2011ui}, a higher spin $dS_4$ duality was 
conjectured involving a 3-dim CFT of anti-commuting $Sp(N)$ (ghost)
scalars.

Certain attempts at generalizing the Ryu-Takayanagi formulation
\cite{Ryu:2006bv,Ryu:2006ef,HRT,HEEreview,HEEreview2} to $dS/CFT$ were
carried out in \cite{Narayan:2015vda,Narayan:2015oka}: while
appropriate real surfaces were found to have vanishing area, the areas
of certain complex codim-2 extremal surfaces (involving an imaginary
bulk time parametrization) were found to have structural resemblance
with entanglement entropy of dual Euclidean CFTs. These end up being
equivalent to analytic continuation from the Ryu-Takayanagi
expressions in $AdS/CFT$.  In $dS_4$ the areas are real and
negative. Towards gaining some insight into whether such a negative
entanglement entropy can at all arise in a field theoretic
calculation, certain 2-dim ghost conformal field theories with
negative central charge were studied as toy models for the replica
calculation in \cite{Narayan:2016xwq}. Specifically certain $c=-2$
ghost-CFTs were focussed upon, where (i) the $SL(2)$ vacuum coincides
with the ghost ground state and (ii) correlation functions are
calculated in the presence of appropriate ghost zero mode insertions
which cancel the background charge inherent in these systems. The
replica formulation via twist operator 2-point correlation functions
then gives the entanglement entropy for a single interval of size $l$
as the usual ${c\over 3}\log {l\over\epsilon}$ behaviour (with
$\epsilon$ the ultraviolet cutoff): this is negative and has various
odd properties as discussed there.  Also studied in
\cite{Narayan:2016xwq} was a toy model of two ghost-spins with a view
to exploring a simple quantum mechanical system with negative norm
states, with the ghost-spin defined as we have reviewed in
sec.~\ref{sec:revi-spins-ghost}. The reduced density matrix obtained
by tracing over one ghost-spin then reveals that positive norm states
give positive von Neumann entropy while negative norm states give
entanglement entropy with a negative real part and a constant
imaginary part.  Overall these are perhaps best regarded as formal
generalizations of the ideas and techniques of the usual notions of
entanglement entropy in more familiar quantum systems.  While a
deeper understanding, if any, of this dual entanglement entropy 
(although consistent with negative central charge) as a probe of $dS/CFT$
remains open, our interest in the present paper has been to study the
resulting object in toy quantum mechanical systems of entangled spins
and ghost-spins towards exploring patterns of quantum
entanglement in systems containing negative norm states that are
expected to arise in systems with a gauge symmetry as mentioned
earlier.

Finally, in light of the present analysis where we have seen that 
generically negative norm states give a complex-valued entanglement 
entropy\footnote{although note that none of these models gives a pure 
imaginary entanglement entropy, as might arise in the case of 
$dS_3/CFT_2$ where the central charge is pure imaginary.}, we
recall that the replica calculation for the $c=-2$ 2-dim ghost CFTs in
\cite{Narayan:2016xwq} recovered only the negative real part, with no
imaginary part. It is interesting to note that an extra phase $(-1)^n$ 
in the reduced density matrix $\rho_A^n\ra (-1)^n\rho_A^n$ in the 
replica theory gives a contribution $S_A\ra S_A - \del_n\log(-1)^n = 
S_A \pm i\pi$ in the $n\ra 1$ limit. We hope to understand this 
and related issues better.

\vspace{15mm}

{\footnotesize \noindent {\bf Acknowledgements:}\ \ It is a pleasure
  to thank Anshuman Maharana and especially Ashoke Sen for several
  useful discussions. KN thanks the hospitality of the String Group,
  HRI, Allahabad, where this work began, and the Organizers of the
  Simons Workshop in Mathematics and Physics, 2016, Simons Center,
  Stony Brook, USA for hospitality while this work was in progress.
  The work of KN is partially supported by a grant to CMI from the Infosys
  Foundation and of DPJ by the DAE project 12-R\&D-HRI-5.02-0303.}

\vspace{3mm}

\appendix

\section{Spin \& ghost-spin:\ \ off-diagonal density matrix}
\label{sec:diag-dens-matr}

We will consider a system of one spin and one ghost-spin here.  A general
state in this set up contains 4 parameters.  In 
sec.~\ref{sec:one-spin-entangled}, we tried to
impose a constraint on these parameters so that the resulting density
matrix was diagonal.

Here we will not demand the diagonal form of the density matrix to
start with.  However, it can always be diagonalised by a change
of basis.  We can then try to find conditions under which the reduced 
density matrix is positive.  Recall the general form of $\rho_A$
in this system is
\begin{equation}
  \label{eq:1}
  \begin{split}
    (\rho_A)^{++} = |\psi^{++}|^2 - |\psi^{+-}|^2\ , & \qquad
(\rho_A)^{+-} = \psi^{++} (\psi^*)^{-+} - \psi^{+-} (\psi^*)^{--}\ ,\nonumber\\
(\rho_A)^{-+} = \psi^{-+} (\psi^*)^{++} - \psi^{--} (\psi^*)^{+-}\ ,& \qquad
(\rho_A)^{--} = |\psi^{-+}|^2 - |\psi^{--}|^2\ .
  \end{split}
\end{equation}
It can be written as 
\begin{align}
  \label{eq:2}
  \rho_A =
  \begin{pmatrix}
     |\psi^{++}|^2 - |\psi^{+-}|^2 & \psi^{++} (\psi^*)^{-+} - \psi^{+-}
     (\psi^*)^{--}\\ 
     \psi^{-+} (\psi^*)^{++} - \psi^{--} (\psi^*)^{+-}&|\psi^{-+}|^2 - |\psi^{--}|^2
  \end{pmatrix}\ .
\end{align}
We can diagonalise this by solving the quadratic equation
\begin{equation}
  \label{eq:3}
  \begin{split}
    0=&\lambda^2 - ( |\psi^{++}|^2 - |\psi^{+-}|^2+|\psi^{-+}|^2 -
  |\psi^{--}|^2) \lambda \\ &+ (|\psi^{++}|^2 - |\psi^{+-}|^2)(|\psi^{-+}|^2 -
  |\psi^{--}|^2)\\ &-(\psi^{++} (\psi^*)^{-+} - \psi^{+-}
     (\psi^*)^{--})(\psi^{-+} (\psi^*)^{++} - \psi^{--} (\psi^*)^{+-})\ .
  \end{split}
\end{equation}
We can use the fact that ${\rm Tr}\rho = \pm 1$ to write
\begin{equation}
  \label{eq:5}
  \begin{split}
    0=&\lambda^2 \pm \lambda + (|\psi^{++}|^2 -
    |\psi^{+-}|^2)(|\psi^{-+}|^2 - 
  |\psi^{--}|^2)\\ &-(\psi^{++} (\psi^*)^{-+} - \psi^{+-}
     (\psi^*)^{--})(\psi^{-+} (\psi^*)^{++} - \psi^{--} (\psi^*)^{+-})\ .
  \end{split}
\end{equation}
The solution to this equation is
\begin{equation}
  \label{eq:4}
  \begin{split}
    \lambda = \pm \frac12 \pm \frac{1}{2}\sqrt{1 + 4(\psi^{++}\psi^{--} -
      \psi^{+-}\psi^{-+})((\psi^*)^{++}(\psi^*)^{--} -
      (\psi^*)^{+-}(\psi^*)^{-+})}]
  \end{split}
\end{equation}
Clearly we need to impose a variety of constraints to ensure that the
eigenvalues are not complex and to ensure that they are positive definite.
First of all, if the trace of the density matrix is $\pm 1$ then we get both
positive (negative) eigenvalues if
\begin{equation}
  \label{eq:6}
 -\frac{1}{4} \leq (\psi^{++}\psi^{--} -
      \psi^{+-}\psi^{-+})((\psi^*)^{++}(\psi^*)^{--} -
      (\psi^*)^{+-}(\psi^*)^{-+}) < 0\ .
\end{equation}
This condition is not satisfied because the quantity we are looking at
is the modulus square of $(\psi^{++}\psi^{--} -\psi^{+-}\psi^{-+})$
which is positive semi-definite and as a result we have one positive
and one negative eigenvalue.

Thus we see that if we do not impose any conditions on the parameters
of the entangled state we seem to get one positive and one negative
eigenvalue of the density matrix.

\section{Tracing over a spin and a ghost-spin}

In sec.~\ref{sec:one-spin-entangled} and sec.~\ref{sec:k+1}, we had 
analysed tracing over the single ghost-spin with the result that 
positive norm states are not correlated with positive entanglement.
One could ask if any alternate mechanism of tracing over a subsector
could make sense in this system.  To explore that
let us now consider two spins entangled with one ghost-spin.  As we
saw above this is similar to the case of one spin and
one ghost-spin.  The minus signs that arise in the
$(\rho_A)^i_k$ components come from the trace over the single ghost-spin
since the spin indices contract with $\delta_{ij}$.  Therefore it
appears that the fate of this system is more in the hands of the
ghost-spin sector than the spin sector.

However suppose we view this system as a single spin entangled with 
an entangled system of one spin and one ghost-spin, and trace over the
latter, \ie\ trace over the entire entangled one spin--one ghost-spin
system.  We can then ask if the reduced density matrix $\rho_A$  
for the single spin looks physically reasonable, or not.  In
particular, does $\rho_A$ possess positivity?

A generic state for two spins and one ghost-spin system and its norm are
\be
|\psi\rangle = \psi^{ij,\al} |ij\rangle |\al\rangle\ ,\qquad
\langle\psi|\psi\rangle = g_{ij} g_{kl} \gamma_{\al\beta}
\psi^{ij,\al}(\psi^*)^{kl,\beta}\ .
\ee
Since we have only one ghost-spin, the norm has just one $\gamma_{ik}$
factor.
For instance, the state 
\be\label{psi21gen}
|\psi\rangle = \psi^{++,+} |++\rangle |+\rangle 
+ \psi^{++,-} |++\rangle |-\rangle + \psi^{--,+} |--\rangle |+\rangle 
+ \psi^{--,-} |--\rangle |-\rangle
\ee
which in general is not a product state has the norm
\be
\langle\psi|\psi\rangle = |\psi^{++,+}|^2 - |\psi^{++,-}|^2
+ |\psi^{--,+}|^2 - |\psi^{--,-}|^2 = \pm 1\ .
\ee

The single spin reduced density matrix obtained by tracing over
the ghost-spin and one spin is
\be\label{rhoA21}
(\rho_A)^{ik} = g_{jl} \gamma_{\al\beta} 
\psi^{ij,\al} (\psi^*)^{kl,\beta} = 
g_{jj} \gamma_{\al\al} \psi^{ij,\al} (\psi^*)^{kj,\al}\ .
\ee
For the choice of entangled state \eqref{psi21gen} we end up getting the
diagonal form of the reduced density matrix
\bea
(\rho_A)^{+,+} = |\psi^{++,+}|^2 - |\psi^{++,-}|^2\ , & \quad &
(\rho_A)^{+,-} = 0\ ,\nonumber\\
(\rho_A)^{-,+} = 0\ , & \quad &
(\rho_A)^{-,-} = |\psi^{--,+}|^2 - |\psi^{--,-}|^2\ .
\eea
As a consequence of the diagonal form of $\rho_A$, $\log\rho_A$ also
has a simple form. Note 
that $tr\rho_A = g_{ik} (\rho_A)^{ik} = (\rho_A)^{+,+} + (\rho_A)^{-,-}$ 
and satisfies $tr\rho_A = tr\rho = \langle\psi|\psi\rangle$. Since 
the remaining spin has positive definite metric $g_{ij}$, we have 
the entanglement entropy
\be
S_A = -g_{ij} (\rho_A\log\rho_A)^{ij} = - (\rho_A\log\rho_A)^{+,+} 
- (\rho_A\log\rho_A)^{-,-}\ .
\ee
Then the entanglement entropy is
\be\label{rhoA21-x}
\begin{split}
  |\psi^{++,+}|^2 - |\psi^{++,-}|^2 \equiv x ,\qquad 
\langle\psi|\psi\rangle = x + (\pm 1-x) ;\\ 
(\rho_A)^{+,+} =  x ,\qquad
(\rho_A)^{-,-} = \pm 1 - x ,\quad \\
S_A = - x\log x - (\pm 1 - x) \log (\pm 1 - x)\ .
\end{split}
\ee
Curiously this structure is similar to the case of one spin entangled
with two ghost-spins.  It would be interesting to relate this system
to a more physical situation to gain insight into this pattern of
entanglement.

To see whether this conclusion survives when we make a different
choice of entangled state, let us consider the most general state
\be\label{psi21mostgen}
\begin{split}
  |\psi\rangle =& \ \psi^{++,+} |++\rangle |+\rangle 
+ \psi^{++,-} |++\rangle |-\rangle + \psi^{+-,+} |+-\rangle |+\rangle
+ \psi^{+-,-} |+-\rangle |-\rangle \\
&\ \ +\ \psi^{-+,+} |-+\rangle |+\rangle + \psi^{-+,-} |-+\rangle
|-\rangle + \psi^{--,+} |--\rangle |+\rangle + \psi^{--,-} |--\rangle
|-\rangle . 
\end{split}
\ee
The norm of this state is
\be
\begin{split}
  \langle\psi|\psi\rangle =& \ |\psi^{++,+}|^2 - |\psi^{++,-}|^2
+ |\psi^{+-,+}|^2 - |\psi^{+-,-}|^2 \\ &\ \ + |\psi^{-+,+}|^2 - |\psi^{-+,-}|^2
+ |\psi^{--,+}|^2 - |\psi^{--,-}|^2
\end{split}
\ee
The reduced density matrix after tracing over a spin and a ghost spin is
\bea\label{rhoA21mostgen}
(\rho_A)^{+,+} &=& |\psi^{++,+}|^2 - |\psi^{++,-}|^2 + |\psi^{+-,+}|^2
- |\psi^{+-,-}|^2\ , \qquad \nonumber\\
(\rho_A)^{+,-} &=& \psi^{++,+}(\psi^*)^{-+,+} - \psi^{++,-}(\psi^*)^{-+,-}
+ \psi^{+-,+}(\psi^*)^{--,+} -  \psi^{+-,-}(\psi^*)^{--,-}\ ,\nonumber\\
(\rho_A)^{-,+} &=& \psi^{-+,+}(\psi^*)^{++,+} - \psi^{-+,-}(\psi^*)^{++,-}
+ \psi^{--,+}(\psi^*)^{+-,+} -  \psi^{--,-}(\psi^*)^{+-,-}\ , \qquad \\
(\rho_A)^{-,-} &=& |\psi^{-+,+}|^2 - |\psi^{-+,-}|^2 
+ |\psi^{--,+}|^2 - |\psi^{--,-}|^2\ .\nonumber
\eea
If we set the off diagonal components of $\rho_A$ to zero then
\begin{equation}
  \label{eq:10}
  \begin{split}
  \psi^{++,+} &= \frac{1}{(\psi^*)^{-+,+}}\left( \psi^{++,-}(\psi^*)^{-+,-}
    - \psi^{+-,+}(\psi^*)^{--,+} +
    \psi^{+-,-}(\psi^*)^{--,-}\right)\\
   \psi^{--,-} &= \frac{1}{(\psi^*)^{+-,-}}\left(\psi^{-+,+}(\psi^*)^{++,+}
   - \psi^{-+,-}(\psi^*)^{++,-} + \psi^{--,+}(\psi^*)^{+-,+}\right) \ .
  \end{split}
\end{equation}
The diagonal components can be rewritten using eq.\eqref{eq:10},
\begin{equation}
  \label{eq:11}
  \begin{split}
    (\rho_A)^{+,+} &=\frac{1}{|\psi^{-+,+}|^2}\bigg[
    |\psi^{++,-}|^2(|\psi^{-+,-}|^2 
      -|\psi^{-+,+}|^2)+ |\psi^{+-,+}|^2(|\psi^{--,+}|^2+|\psi^{-+,+}|^2) \\
      &+|\psi^{+-,-}|^2(|\psi^{--,-}|^2 - |\psi^{-+,+}|^2) +\big\{
      \psi^{++,-}(\psi^*)^{-+,-}\psi^{--,-}(\psi^*)^{+-,-} \\
      &- \psi^{++,-}(\psi^*)^{-+,-}\psi^{--,+}(\psi^*)^{+-,+} -
      \psi^{+-,+}(\psi^*)^{--,+}\psi^{--,-}(\psi^*)^{+-,-} + c.c.
      \big\} \bigg]\\
      (\rho_A)^{-,-}
      &=-\frac{1}{|\psi^{+-,-}|^2}\bigg[|\psi^{-+,+}|^2(|\psi^{++,+}|^2
      -|\psi^{+-,-}|^2) +
      |\psi^{-+,-}|^2(|\psi^{++,-}|^2+|\psi^{+-,-}|^2)  \\
      &+|\psi^{--,+}|^2(|\psi^{+-,+}|^2-|\psi^{+-,-}|^2)
      +\big\{\psi^{-+,+}(\psi^*)^{++,+}\psi^{+-,+}(\psi^*)^{--,+}\\  
      &-\psi^{-+,+}(\psi^*)^{++,+}\psi^{++,-}(\psi^*)^{-+,-}  -
      \psi^{-+,-}(\psi^*)^{++,-}\psi^{+-,+}(\psi^*)^{--,+}+c.c.\big\}
      \bigg] \ .
  \end{split}
\end{equation}
The form of the diagonal components is rich enough to allow various
possibilities, which clearly include cases where we get positive
entanglement entropy.  For example, if we demand that each term in
$(\rho_A)^{+,+}$ is positive definite then we need to impose three
conditions.  Two of which are 
\begin{equation}
  \label{eq:12}
    |\psi^{-+,-}|^2 \ge |\psi^{-+,+}|^2,\qquad |\psi^{--,-}|^2 \ge
    |\psi^{-+,+}|^2\ ,
\end{equation}
and the third condition puts imposes the positivity condition on the
curly bracket term in the expression for $(\rho_A)^{+,+}$ in
eq.\eqref{eq:11}.  Similarly demanding that $(\rho_A)^{-,-}$ is
negative definite gives following conditions
\begin{equation}
  \label{eq:13}
   |\psi^{++,+}|^2 \ge |\psi^{+-,-}|^2,\qquad |\psi^{+-,+}|^2 \ge
    |\psi^{+-,-}|^2\ ,
\end{equation}
and a positivity constraint on the
curly bracket term in the expression for $(\rho_A)^{-,-}$ in
eq.\eqref{eq:11}.


\begin{thebibliography}{} 

\footnotesize{

\bibitem{AreaLaw1}
  L.~Bombelli, R.~K.~Koul, J.~Lee and R.~D.~Sorkin,
  ``A Quantum Source of Entropy for Black Holes,''
  \pr{D}{34}{1986}{373}.


\bibitem{AreaLaw2}
   M.~Srednicki,
   ``Entropy and area,''
   \prl{71}{1993}{666} [\arXivid{hep-th/9303048}].

\bibitem{Holzhey:1994we} 
  C.~Holzhey, F.~Larsen and F.~Wilczek,
  ``Geometric and renormalized entropy in conformal field theory,''
  \npb{B}{424}{1994}{443}
  [\arXivid{hep-th/9403108}].

\bibitem{Vidal:2002rm} 
  G.~Vidal, J.~I.~Latorre, E.~Rico and A.~Kitaev,
  ``Entanglement in quantum critical phenomena,''
  \prl{90}{2003}{227902}
  [\arXivid{quant-ph/0211074}].

\bibitem{Latorre:2003kg} 
  J.~I.~Latorre, E.~Rico and G.~Vidal,
  ``Ground state entanglement in quantum spin chains,''
  Quant.\ Inf.\ Comput.\  {\bf 4}, 48 (2004)
  [\arXivid{quant-ph/0304098}].

\bibitem{Calabrese:2004eu} 
  P.~Calabrese and J.~L.~Cardy,
  ``Entanglement entropy and quantum field theory,''
  J.\ Stat.\ Mech.\  {\bf 0406}, P06002 (2004)
  [\arXivid{hep-th/0405152}].

\bibitem{Horodecki:2009zz} 
  R.~Horodecki, P.~Horodecki, M.~Horodecki and K.~Horodecki,
  ``Quantum entanglement,''
  \rmp{81}{2009}{865}
  [\arXivid{quant-ph/0702225}].

\bibitem{Eisert:2008ur} 
  J.~Eisert, M.~Cramer and M.~B.~Plenio,
  ``Area laws for the entanglement entropy - a review,''
  \rmp{82}{2010}{277}
  [\arXivid{0808.3773}[quant-ph]].

\bibitem{Calabrese:2009qy} 
  P.~Calabrese and J.~Cardy,
  ``Entanglement entropy and conformal field theory,''
  J.\ Phys.\ A {\bf 42}, 504005 (2009)
  doi:10.1088/1751-8113/42/50/504005
  [\arXivid{0905.4013}[cond-mat.stat-mech]].

\bibitem{Casini:2009sr} 
  H.~Casini and M.~Huerta,
  ``Entanglement entropy in free quantum field theory,''
  J.\ Phys.\ A {\bf 42}, 504007 (2009)
  doi:10.1088/1751-8113/42/50/504007
  [\arXivid{0905.2562}[hep-th]].

\bibitem{Maldacena:1997re}
  J.~M.~Maldacena,
  ``The large N limit of superconformal field theories and supergravity,''
  \atmp{2}{1998}{231}
  [Int.\ J.\ Theor.\ Phys.\  {\bf 38}, 1113 (1999)]
  [\arXivid{hep-th/9711200}].

\bibitem{Gubser:1998bc}
  S.~S.~Gubser, I.~R.~Klebanov and A.~M.~Polyakov,
  ``Gauge theory correlators from non-critical string theory,''
  \pl{B}{428}{1998}{105}
  [\arXivid{hep-th/9802109}].

\bibitem{Witten:1998qj}
  E.~Witten,
  ``Anti-de Sitter space and holography,''
  \atmp{2}{1998}{253}
  [\arXivid{hep-th/9802150}].

\bibitem{Aharony:1999ti}
  O.~Aharony, S.~S.~Gubser, J.~M.~Maldacena, H.~Ooguri and Y.~Oz,
  ``Large N field theories, string theory and gravity,''
  \phyrep{323}{2000}{183}
  [\arXivid{hep-th/9905111}].

\bibitem{Ryu:2006bv} 
  S.~Ryu and T.~Takayanagi,
  ``Holographic derivation of entanglement entropy from AdS/CFT,''
  \prl{96}{2006}{181602}
  [\arXivid{hep-th/0603001}].

\bibitem{Ryu:2006ef} 
  S.~Ryu and T.~Takayanagi,
  ``Aspects of Holographic Entanglement Entropy,''
  \jhep{08}{2006}{045}
  [\arXivid{hep-th/0605073}].

\bibitem{HRT} 
V.~E.~Hubeny, M.~Rangamani and T.~Takayanagi,
``A Covariant holographic entanglement entropy proposal,'' 
\jhep{07}{2007}{062}  [\arXivid{0705.0016}[hep-th]].
 
\bibitem{HEEreview}
  T.~Nishioka, S.~Ryu and T.~Takayanagi,
  ``Holographic Entanglement Entropy: An Overview,''
  J.\ Phys.\ A {\bf 42}, 504008 (2009)
  doi:10.1088/1751-8113/42/50/504008
  [\arXivid{0905.0932}[hep-th]].

\bibitem{HEEreview2}
T.~Takayanagi,
  ``Entanglement Entropy from a Holographic Viewpoint,''
  \cqg{29}{2012}{153001}  [\arXivid{1204.2450}[gr-qc]].

\bibitem{Narayan:2016xwq} 
  K.~Narayan,
  ``On $dS_4$ extremal surfaces and entanglement entropy in some ghost CFTs,''
  \pr{D}{94}{2016}{046001}
  [\arXivid{1602.06505}[hep-th]].

\bibitem{polchinskiTextBk}
J.~Polchinski, String Theory, Vol. 1,2. 
Cambridge University Press (1998).

\bibitem{Casini:2013rba} 
  H.~Casini, M.~Huerta and J.~A.~Rosabal,
  ``Remarks on entanglement entropy for gauge fields,''
  \pr{D}{89}{2014}{085012}
  [\arXivid{1312.1183}[hep-th]].

\bibitem{Ghosh:2015iwa} 
  S.~Ghosh, R.~M.~Soni and S.~P.~Trivedi,
  ``On The Entanglement Entropy For Gauge Theories,''
  \jhep{09}{2015}{069}
  doi:10.1007/JHEP09(2015)069
  [\arXivid{1501.02593}[hep-th]].

\bibitem{Soni:2015yga} 
  R.~M.~Soni and S.~P.~Trivedi,
  ``Aspects of Entanglement Entropy for Gauge Theories,''
  \jhep{01}{2016}{136}
  doi:10.1007/JHEP01(2016)136
  [\arXivid{1510.07455}[hep-th]].

\bibitem{Casini:2015dsg} 
  H.~Casini and M.~Huerta,
  ``Entanglement entropy of a Maxwell field on the sphere,''
  \pr{D}{93}{2016}{105031}
  [\arXivid{1512.06182}[hep-th]].

\bibitem{Soni:2016ogt} 
  R.~M.~Soni and S.~P.~Trivedi,
  ``Entanglement Entropy in (3+1)-d Free $U(1)$ Gauge Theory,''
  [\arXivid{1608.00353}[hep-th]].

\bibitem{Kabat:1995eq} 
  D.~N.~Kabat,
  ``Black hole entropy and entropy of entanglement,''
  Nucl.\ Phys.\ B {\bf 453}, 281 (1995)
  doi:10.1016/0550-3213(95)00443-V
  [\arXivid{hep-th/9503016}].

\bibitem{Donnelly:2015hxa} 
  W.~Donnelly and A.~C.~Wall,
  ``Geometric entropy and edge modes of the electromagnetic field,''
  Phys.\ Rev.\ D {\bf 94}, no. 10, 104053 (2016)
  doi:10.1103/PhysRevD.94.104053
  [\arXivid{1506.05792} [hep-th]].

\bibitem{Harlow:2015lma} 
  D.~Harlow,
  ``Wormholes, Emergent Gauge Fields, and the Weak Gravity Conjecture,''
  JHEP {\bf 1601}, 122 (2016)
  doi:10.1007/JHEP01(2016)122
  [\arXivid{1510.07911} [hep-th]].

\bibitem{Jatkar:2017jwz} 
  D.~P.~Jatkar and K.~Narayan,
  ``Ghost-spin chains, entanglement and $bc$-ghost CFTs,''
  [\arXivid{1706.06828} [hep-th]].

\bibitem{Strominger:2001pn} 
  A.~Strominger,
  ``The dS / CFT correspondence,''
  \jhep{10}{2001}{034}
  [\arXivid{hep-th/0106113}].

\bibitem{Witten:2001kn} 
  E.~Witten,
  ``Quantum gravity in de Sitter space,''
  [\arXivid{hep-th/0106109}].

\bibitem{Maldacena:2002vr}
  J.~M.~Maldacena,
  ``Non-Gaussian features of primordial fluctuations in single field inflationary models,''
  \jhep{05}{2003}{013}
  [\arXivid{astro-ph/0210603}].

\bibitem{Anninos:2011ui} 
  D.~Anninos, T.~Hartman and A.~Strominger,
  ``Higher Spin Realization of the dS/CFT Correspondence,''
  [\arXivid{1108.5735}[hep-th]].

\bibitem{Narayan:2015vda} 
  K.~Narayan,
  ``de Sitter extremal surfaces,''
  \pr{D}{91}{2015}{126011}
  [\arXivid{1501.03019}[hep-th]].

\bibitem{Narayan:2015oka} 
  K.~Narayan,
  ``de Sitter space and extremal surfaces for spheres,''
  \pl{B}{753}{2016}{308}
  doi:10.1016/j.physletb.2015.12.019
  [\arXivid{1504.07430}[hep-th]].
} 
\end{thebibliography}
\end{document}